\documentclass[acmsmall]{acmart}

\usepackage{float, graphicx}
\usepackage{lipsum}
\usepackage{multirow}
\usepackage{tabularx}
\usepackage{csquotes}
\usepackage{dblfloatfix}
\usepackage{natbib}
\usepackage{bm}
\usepackage{amsmath,scalerel}
\usepackage{xfrac}
\usepackage{bbm}
\usepackage{graphicx}
\usepackage{subfig}
\usepackage{algpseudocode}
\usepackage{algorithm}
\usepackage{url}
\AtBeginDocument{%
  }

\begin{document}


\newcommand{\modelplain}{ImaGGen}
\newcommand{\model}{\textbf{\modelplain}}

\title{\modelplain: Zero-Shot Generation of Co-Speech Semantic Gestures Grounded in Language and Image Input}

\author{Hendric Voss}
\email{hvoss@techfak.uni-bielefeld.de}
\orcid{0009-0003-3646-7702}
\affiliation{%
  \institution{Social Cognitive Systems Group, Bielefeld University}
  \streetaddress{Universitätsstraße 25}
  \country{Germany}
}

\author{Stefan Kopp}
\email{skopp@techfak.uni-bielefeld.de}
\orcid{0000-0002-4047-9277}
\affiliation{%
  \institution{Social Cognitive Systems Group, Bielefeld University}
  \streetaddress{Universitätsstraße 25}
  \country{Germany}
}

\ccsdesc[500]{Human-centered computing~Interactive systems and tools}
\ccsdesc[500]{Computing methodologies~Machine learning}
\ccsdesc[500]{Human-centered computing~Interaction devices}
\ccsdesc[500]{Computing methodologies~Real-time simulation}

\keywords{machine learning; deep learning; co-speech gesture generation; semantic gestures; multimodal data; communicative efficacy, iconic gestures, deictic gestures}




\begin{abstract}
Human communication combines speech with expressive nonverbal cues such as hand gestures that serve manifold communicative functions. Yet, current generative AI-based gesture generation approaches are, for the most part, restricted to simple, repetitive beat gestures that accompany the rhythm of speaking but do not contribute to communicating semantic meaning. This paper tackles a core challenge in co-speech gesture synthesis: generating iconic or deictic gestures that are semantically coherent with a verbal utterance. Our basic assumption is that such gestures cannot be derived from language input alone, which inherently lacks the visual meaning that is often carried autonomously by gestures. 
We therefore introduce a zero-shot system that generates gestures from a given language input and additionally is informed by imagistic input, without manual annotation or human intervention.
Our method integrates an image analysis pipeline that extracts key object properties such as shape, symmetry, and alignment, together with a semantic matching module that links these visual details to spoken text. An inverse kinematics engine then synthesizes iconic and deictic gestures and combines them with co-generated natural beat gestures for coherent multimodal communication.
A comprehensive user study demonstrates the effectiveness of our approach. In scenarios where speech alone was ambiguous, gestures generated by our system significantly improved participants' ability to identify object properties, confirming their interpretability and communicative value. While challenges remain in representing complex shapes, our results highlight the importance of context-aware semantic gestures for creating expressive and collaborative virtual agents or avatars, marking a substantial step forward towards efficient and robust, embodied human-agent interaction. More information and example videos are available here: \url{https://review-anon-io.github.io/ImaGGen.github.io/}
\end{abstract}

\begin{teaserfigure}
  \includegraphics[width=\textwidth]{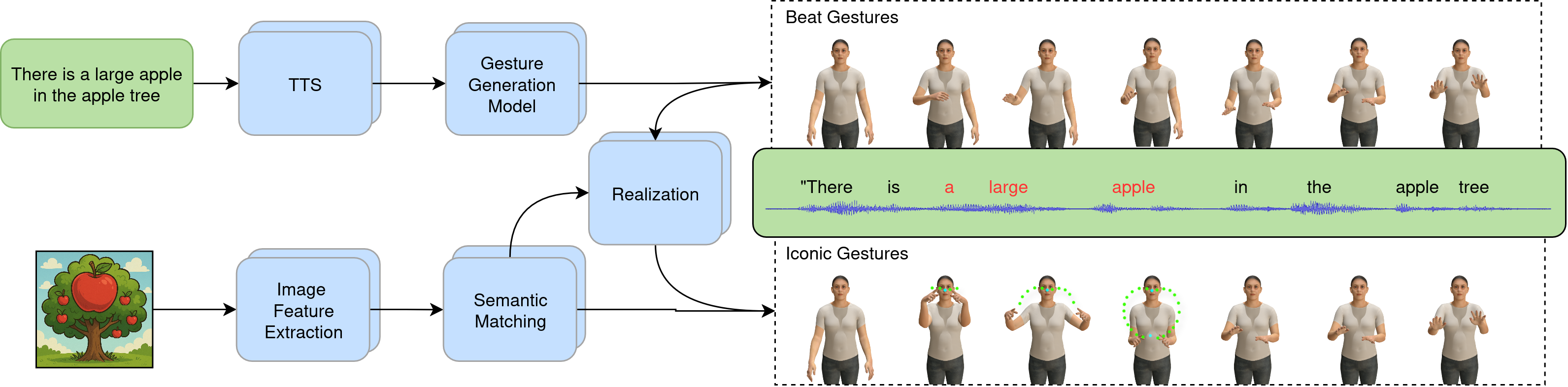}
  \caption{Overview of the \model{} model to create and integrate semantic (iconic and deictic) gestures and beat gestures from any given speech and image input without human supervision.}
  \Description{A diagram illustrates a system that converts spoken language and images into sign language gestures. At the top left, a sentence reads: "There is a large apple in the apple tree." This goes into a Text-to-Speech module, which then feeds into a Gesture Generation Model. Below that, an image of an apple tree with a large red apple is shown, which is processed through an Image Feature Extraction module and then passed to a Semantic Matching module. Both gesture and iconic inputs converge into a Realization module. To the right, two sequences of a woman performing sign language gestures are shown. The upper row aligns the gestures with the spoken sentence, showing synchronization with the audio waveform and highlighting words like "a," "large," and "apple" in red. The bottom row highlights iconic gestures for the word "apple," showing specific hand motions representing the concept visually. The overall process demonstrates how both language and visual content contribute to generating accurate and expressive sign language.}
  \label{fig:teaser}
\end{teaserfigure}

\maketitle


\section{Introduction}

Human communication is inherently multimodal, combining spoken language with a rich array of nonverbal signals \cite{frohlich2019multimodal}. Among these, hand gestures hold a particularly prominent role. Far from being mere accompaniments to speech, gestures contribute actively to the conveyance of information and interpersonal understanding \cite{goldin2013gesture,wagner2014}. They serve as visual representations of concepts, actions, or relationships, often bridging the gap between what can be easily verbalized and what is challenging to express in words \cite{kang2016hands, goldin2010action}. This multimodal integration reflects the deeply embodied nature of human cognition and communication, where gestures and language work in tandem to convey complex ideas and facilitate mutual comprehension \cite{iverson1999hand, quek2002multimodal}.

Within the spectrum of co-speech gestures, iconic gestures are especially significant. These gestures depict spatial aspects of a referent through forms or motion, providing visual cues that relate directly to the spoken content \cite{krauss1998we}. Iconic gestures function as a central component of human communication, influencing language processing during both speech production and comprehension \cite{beattie2001experimental, Kopp2005TheSS}. They are employed for a variety of communicative purposes. Speakers often use iconic gestures to emphasize critical information, thereby ensuring its prominence in the discourse \cite{10.1145/3477322.3477330}. They can also rapidly convey additional details about shapes, movements, or spatial relationships that would require lengthy verbal descriptions \cite{capone2004gesture}. In many cases, iconic gestures express meanings that are difficult or even impossible to articulate solely through speech, such as abstract actions, unfamiliar shapes, or nuanced physical transformations \cite{beattie1999mapping}.

Research in cognitive science has long emphasized that iconic gestures emerge naturally as part of speech production rather than as secondary add-ons. Foundational work by \citet{Kendon1981GesticulationAS} and \citet{McNeill1992HandMind} proposed that gesture and language originate from a shared conceptual source, allowing gestures to externalize aspects of mental imagery and spatial reasoning.
Experimental studies further demonstrate that gestures support lexical access and fluency during speech \cite{Rauscher1996GestureLexical}, and their absence can disrupt conceptual planning when speakers describe visuospatial content.
For listeners, iconic gestures enhance comprehension, inference-making, and memory \cite{Kelly1999Pragmatic, Hostetter2011Meta}. Neurocognitive evidence corroborates these findings, showing that gesture-speech mismatches evoke stronger brain responses and processing delays, suggesting that the two modalities are tightly integrated at early stages of comprehension \cite{Ozyurek2007ERP, Kelly2010StroopGesture}.
Listeners even form expectations about gestures based on speech context, with missing or incongruent gestures eliciting increased processing load \cite{Wu2010MissingGestures}. These results highlight that gestures are not merely expressive but form an integral and expected channel in multimodal communication.

The importance of iconic gestures also becomes evident early on in human development. Children as young as 21 months have been shown to invent novel iconic gestures to communicate information to others \cite{behne2014young}. Moreover, these young learners can interpret iconic gestures and use them to infer previously unknown actions \cite{novack2015learning}. This early ability underscores the cognitive and communicative significance of iconic gestures as a powerful and important tool for conveying information. Notably, this ability extends beyond human speakers: listeners can interpret iconic gestures performed by humanoid robots as effectively as those made by humans \cite{Bremner2016RobotAvatars}. This suggests that gesture comprehension relies on general cognitive mechanisms, reinforcing their potential in artificial agents and virtual environments.

Despite their central role in natural communication, iconic gestures have not been satisfactorily covered in automatic co-speech gesture generation in the field of Computer Graphics/Animation or virtual agents \cite{wolfert2022review}. While early work has proposed planning or rule-based approaches to derive iconic gesture features and animate them procedurally \cite{bergmann2009gnetic,tepper2004content}, current systems are based on generative AI models and have shifted focus to generating human-like gestural motion for a given language input (text or speech). The resulting behavior for the most part consists of beat gestures that are temporally aligned with prosodic features of speech but convey very little or no semantic information at all \cite{nyatsanga2023comprehensive}. Only a few approaches have recently attempted to incorporate semantic cues in order to approximate human-like iconic gestures \cite{mughal2024convofusion, mughal2025retrieving}. They, however, have achieved only very limited success so far as they struggle with iconic gestures being highly variable, complex, and context-dependent, requiring sensitivity to the speaker's underlying communicative intent and how (parts of) it figures in the semantic or pragmatic properties of the utterance \cite{bergmann2009gnetic}.


Incorporating iconic gestures into co-speech gesture generation systems, hence requires models that consider not only verbal content but also regard factors such as multimodal communicative intent, prior conversational history, or knowledge about the interaction partner \cite{bergmann2012production, masson2017we, tepper2004content}. 
Although prior work has made significant progress in generating beat gestures and, to a much lesser extent, semantic gestures, there remains a clear gap in the ability to produce truly context-aware iconic or deictic gestures solely from a given language input, and to combine those with natural beat gestures in an automatic, scalable, and convincing fashion \cite{nyatsanga2023comprehensive,shimoyama2022visture}. This limitation becomes especially clear in scenarios where gestures must convey properties such as object shape, size, or spatial arrangements that cannot be adequately expressed in speech and hence are not directly accessible from verbal input alone \cite{hassemer2018decoding}. Overcoming this limitation requires methods that integrate multimodal information, including text, speech, imagery, and timing cues. Earlier work has required users to manually pre-specify which gestures should be performed in a visual or verbal context \cite{kopp2004,le2012common, kadono2016automaticJSAI, nihei2019determining} or has tried to automatically derive gestural features from additional representations of imagistic meaning \cite{bergmann2009gnetic}. In sum, however, all existing systems have struggled to reconcile demands for naturalness and human-likeness of gestural motions with the requirement to generate believable and informative gestures that accurately fit both the semantic content of speech and the visual characteristics of an entity to be referred to.

To address this gap, we present \modelplain, a novel approach for the automatic generation of beat, iconic, and deictic gestures without the need for manual annotation or human intervention. As shown in Fig.~\ref{fig:model_architecture}, our model operates on both language and image input and produces semantic gestures that comprise beat, deictic, and iconic components, all synchronized with either recorded or synthesized speech. To the best of our knowledge, \modelplain is the first fully automatic zero-shot system capable of extracting iconic and deictic information from images and generating corresponding gestures on a virtual agent. Our contributions are as follows: First, we introduce an image extraction pipeline that identifies and extracts relevant object information from a given image and retrieves advanced structural properties, such as symmetry, geometric primitives, and alignment for each object. This is crucial, as the combined information can be later used to derive meaningful iconic and deictic gestures that accurately represent the visual characteristics of the object. Second, we propose a semantic matching approach that aligns speech content with visual information by extracting meaningful text and audio timing cues and combining these with image-derived features to generate appropriate iconic and deictic gestures. Third, we develop an inverse kinematics approach based on the work of \citet{Voss2025RealTimeIK}, which synthesizes these gestures in a believable and informative manner while layering them on top of beat gestures that are automatically generated using a state-of-the-art gesture synthesis system. Finally, we quantitatively evaluate our system through a 2x2 human evaluation study that assesses both the recognition accuracy of the generated gestures as well as participants' self-reported confidence in their interpretations. In addition, we employ the ASAQ survey to measure effects on user perception of the virtual agent along seven dimensions: Believability, Performance, Likeability, Acceptance, Enjoyability, Engagement, and Coherence \cite{FITRIANIE2025103482}.

\section{Related Work}

\subsection{Co-Speech Gesture Generation}

The automatic generation of co-speech gestures has long been a central challenge in the development of embodied agents and virtual characters \cite{kurokawa_gesture_1992, cassell1994animated}. Early systems relied on rule-based techniques or manually authored scripts that handcrafted mappings between linguistic content and gesture templates. For instance, \citet{cassell1994animated} introduced an agent architecture where a dialogue planner determined facial expressions, gaze, and hand gestures using rule-based generators. Another prominent example is the \textit{Behavior Expression Animation Toolkit} (BEAT), which linked textual annotations to predefined gesture templates based on discourse function \cite{cassell2001beat}. \citet{stone2004speaking} proposed one of the first data-driven methods by recombining recorded gesture motion samples based on perceptual alignment with speech, although gesture synthesis remained template-based. \citet{hartmann2006implementing} presented \textit{Greta}, a virtual agent using utterance-level communicative functions to select matching gestures with parametric control over their expressiveness. \citet{kopp2004} presented a model to plan and procedurally generate co-speech gesture animations from specifications of desired form features, paving the way for the \textit{Behavior Markup Language} (BML) that standardized the specification of multimodal behavior across virtual agents, enabling synchronized gestures, gaze, and facial expressions \cite{kopp_towards_2006, vilhjalmsson_behavior_2007}.

A significant shift occurred in the 2010s with the introduction of machine learning for gesture generation. \citet{huang2014learning} marked an early transition by using dynamic Bayesian networks (DBNs) to learn temporal alignments between speech, gaze, and gestures from human narrators. This data-driven approach enabled a storytelling robot to mimic human-like multimodal coordination.

With the rise of deep learning, the field moved toward neural architectures trained on large-scale gesture-speech datasets. \citet{yoon2019robots} trained an encoder-decoder model on TED talks to generate gesture sequences from input speech text, producing iconic and beat gestures rated as natural and contextually appropriate. This work was extended in \citet{yoon2020speech} to include audio and a speaker-style embedding, using a GAN-based model to improve synchronization and introduce a Fréchet Gesture Distance metric for evaluation.

Multimodal approaches became more prominent as systems began combining audio and text features to drive gesture generation. For example, \citet{kucherenko2020gesticulator} introduced \textit{Gesticulator}, a framework that fused speech audio and textual features to generate expressive beat and iconic gestures. Around the same time, researchers began focusing on the generative nature of the task, acknowledging that a given speech input could map to multiple valid gesture outputs. \citet{li2021audio2gestures} addressed this by designing a conditional variational autoencoder (VAE) that disentangles shared and gesture-specific latent spaces, enabling diverse and realistic gesture generation.

Recent work has embraced generative modeling as a way to further improve diversity and realism. Diffusion models, in particular, have gained traction. \citet{zhu2023taming} introduced \textit{DiffGesture}, a diffusion-based model that denoises random pose sequences conditioned on speech audio using a transformer architecture and a gesture stabilizer for smooth temporal coherence. \citet{shen2025tedculture} extended this approach to multilingual settings, introducing the TED-Culture dataset and deploying their diffusion-based model on a real-time NAO robot across multiple languages.

Several other recent systems combine generative and discriminative objectives to improve motion quality and speaker style matching \cite{ferstl2019multi, habibie2021learning, liu2022beat}. Techniques such as GANs \cite{wu2021modeling, wu2021probabilistic}, normalizing flows \cite{alexanderson2020style}, VAEs \cite{ghorbani2022zeroeggs}, and VQ-VAEs \cite{yazdian2022gesture2vec, voss2023aq} have all been explored. Some methods even combine multiple generative strategies, such as the hybrid VAE-flow model proposed by \citet{taylor2021speech} or transformer-diffusion hybrids like those of \citet{ng2024audio2photoreal}.

The field has also moved toward richer multimodal generation. Systems now often generate gestures along with facial animation \cite{ng2024audio2photoreal} or synchronized speech audio \cite{wang2021integrated, mehta2023diff, mehta2024unified, mehta2024fake, zhang2025fasttalker}. This reflects broader trends in generative AI towards integrated multimodal synthesis.

\subsection{Automatic Generation of Iconic Gestures}

Early systems for automatic iconic gesture generation laid the foundation for aligning gestures with speech. As one of the earliest methods, BEAT \cite{cassell2001beat} automatically synchronized co-speech gestures with speech prosody using handcrafted rules. Although primarily targeting beat gestures, BEAT also supported the generation of simple rule-based iconic gestures, revealing the potential of prosody-aligned iconic gesture production.

Subsequent work began focusing on generating iconic gestures by deriving their form features from visual and spatial semantics. \citet{tepper2004content} developed an embodied agent that produced iconic gestures for spatial descriptions by integrating gesture planning with a language planning module. \citet{bergmann2009increasing} employed a Bayesian network, called the Gesture Formulator, trained on a gesture corpus to determine when to gesture, what technique to use (e.g., tracing or drawing), and which trajectories or handshapes are appropriate given an intended visual meaning and discourse context. This allowed an agent to compose gesture feature matrices that create iconicity. Building on this, \citet{bergmann2009gnetic} proposed \textit{GNetIc}, which used a Bayesian decision network trained on human speaker data to combine data-driven feature selection with rule-based inference of gesture form features. GNetIc employed perceptual (visual) and linguistic information within the graphical model, enabling autonomous decisions about when and how to produce gestures that express visuospatial meaning.

\citet{le2012common} introduced a modular architecture for gesture generation applicable to both virtual and physical agents. Their system mapped communicative intentions to symbolic gesture templates using an intent and behavior planner (FML/BML), which were then realized as motor actions. A key element was a gesture lexicon that linked high-level meanings to gesture forms. By decoupling gesture specification from embodiment, the same iconic gesture definitions could be applied across humanoid robots and virtual avatars.

Other work explored visual grounding through direct image-based methods. \citet{kadono2016automaticJSAI} developed a system that generates drawing-like iconic gestures from object images. After classifying an object’s shape (e.g., circle or rectangle), the system selected corresponding motion primitives from a predefined dictionary and synchronized them with synthesized speech. These gestures effectively illustrated the shape and functional role of the object being described, demonstrating a template-based visual gesture mapping approach.

Extending visual grounding to mental imagery, \citet{nihei2019determining} trained a deep neural network to predict iconic gestures based on object images. Participants were asked to imagine and sketch objects like a "pen," and these sketches were averaged into canonical visual forms. The model then classified these into one of seven predefined gesture types (e.g., vertical line, circle), achieving around 62\% accuracy. This showed that statistical features of visual or mental imagery could be used to guide semantically aligned gesture generation.

More recently, \citet{shimoyama2022visture} introduced \textit{VISTURE}, a system that generates robot speech and gestures directly from video input. Using YOLOv5 for object detection and shape analysis, the system identified relevant scene elements to describe and produce spoken output along with synchronized form-descriptive gestures such as contour tracing. This approach demonstrated an integrated pipeline for visual analysis and gesture production in scene explanation tasks.

Collectively, these systems span a range of strategies, including rule-based planning, gesture templates, probabilistic graphical models, and deep learning. Despite presenting a proof of concept for image-based co-speech gesture generation, however, they were all restricted to a rather limited range of iconic gestures that, in addition, are not embedded in a natural flow of multimodal utterances with co-verbal beat gestures. Further, none of these systems has released open-source implementations, limiting the possibility to verify results, compare methods, or build on them for future research.

\subsection{Inverse Kinematics for Virtual Humans}
Inverse kinematics (IK) for virtual humans has evolved into a range of analytical, numerical, and learning-based approaches. \citet{Tolani2000IK} introduced a hybrid solver that combines analytic solutions for certain limb segments with numerical refinement for remaining joints. This approach offers more efficient and stable performance than purely Jacobian or optimization-based techniques.
Building on iterative methods, \citet{Aristidou2011FABRIK} proposed \textit{FABRIK}, an efficient IK algorithm that repositions joints through forward and backward passes along straight lines to the target. This method converges quickly, handles joint constraints, and generates realistic motion, making it ideal for real-time applications. More recent efforts have sought closed-form solutions. \citet{Pfurner2016ClosedFormIK} presented an analytic IK solver for human-like arms that yields a one-parameter family of solutions, preserving redundancy for secondary objectives such as singularity avoidance or joint-limit control. Related works target virtual reality applications. \citet{Parger2018UpperBodyIK} introduced a lightweight IK algorithm that uses only head and hand trackers to infer natural upper-body motion. Their heuristics for elbow placement lead to more believable avatar embodiment, outperforming both disembodied hands and full MoCap systems in user studies.

Learning-based IK has emerged as a powerful alternative. \citet{Phaniteja2018StableIK} trained a deep reinforcement learning (RL) agent to perform IK while maintaining balance on a high-DOF humanoid. The agent learns stable reaching behaviors with high accuracy. Extending this, \citet{Singamaneni2018MultiGoalIK} trained a policy to control both arms simultaneously, incorporating torso movement and ensuring double-support balance. 

Hybrid approaches that combine analytic and neural techniques have also shown promise. \citet{Li2021HybrIK} proposed \textit{HybrIK}, which analytically solves swing rotation and predicts twist angles with a neural network. This decomposition improves interpretability and accuracy over purely learned or optimization-based systems.
Traditional numerical solvers remain relevant. \citet{Malik2022DRLIK} discussed the limitations of Jacobian-based IK and highlighted modern constrained solvers like \textit{TRAC-IK}, which better handle joint limits and singularities through optimization.

Newer neural models are designed to leverage skeletal structure. \citet{Agrawal2023SKELIK} presented \textit{SKEL-IK}, a transformer conditioned on sparse joint information and the skeleton graph. This structure-aware design produces high-fidelity full-body poses with minimal inputs. \citet{Li2023NIKI} proposed \textit{NIKI}, an invertible neural network that integrates an analytic twist-and-swing decomposition. Its bidirectional architecture improves robustness to occlusion and yields accurate, interpretable predictions.
\cite{Davoodnia2024SkelFormer} presented \textit{SkelFormer}, which decoupled 3D joint triangulation and pose estimation, using a skeletal transformer to infer full-body pose from multi-view input. The system achieves high accuracy and robustness in noisy environments. In specialized contexts such as medical motion capture, \citet{Korol2024MedicalIK} showed that pre-trained neural networks can outperform traditional solvers. Their LSTM-based model for a 3-DOF human arm is robust to noise and runs faster than real-time, making it suitable for rehabilitation and clinical use.

Most recently, \citet{Voss2025RealTimeIK} has introduced a differentiable, real-time IK solver running entirely within TensorFlow. By treating both forward and inverse kinematics as an optimization problem akin to machine learning, this method enables gradient-based optimization through automatic differentiation and efficiently handles multiple end-effector constraints or joint limits, outperforming classical solvers in convergence, speed, and success rate on complex skeletons. Since this approach supports the real-time generation of complex full-body and hand gestures for virtual humans, we will build on it in the model presented here.

\begin{figure}[tb]
    \centering
    \includegraphics[width=1.0\linewidth]{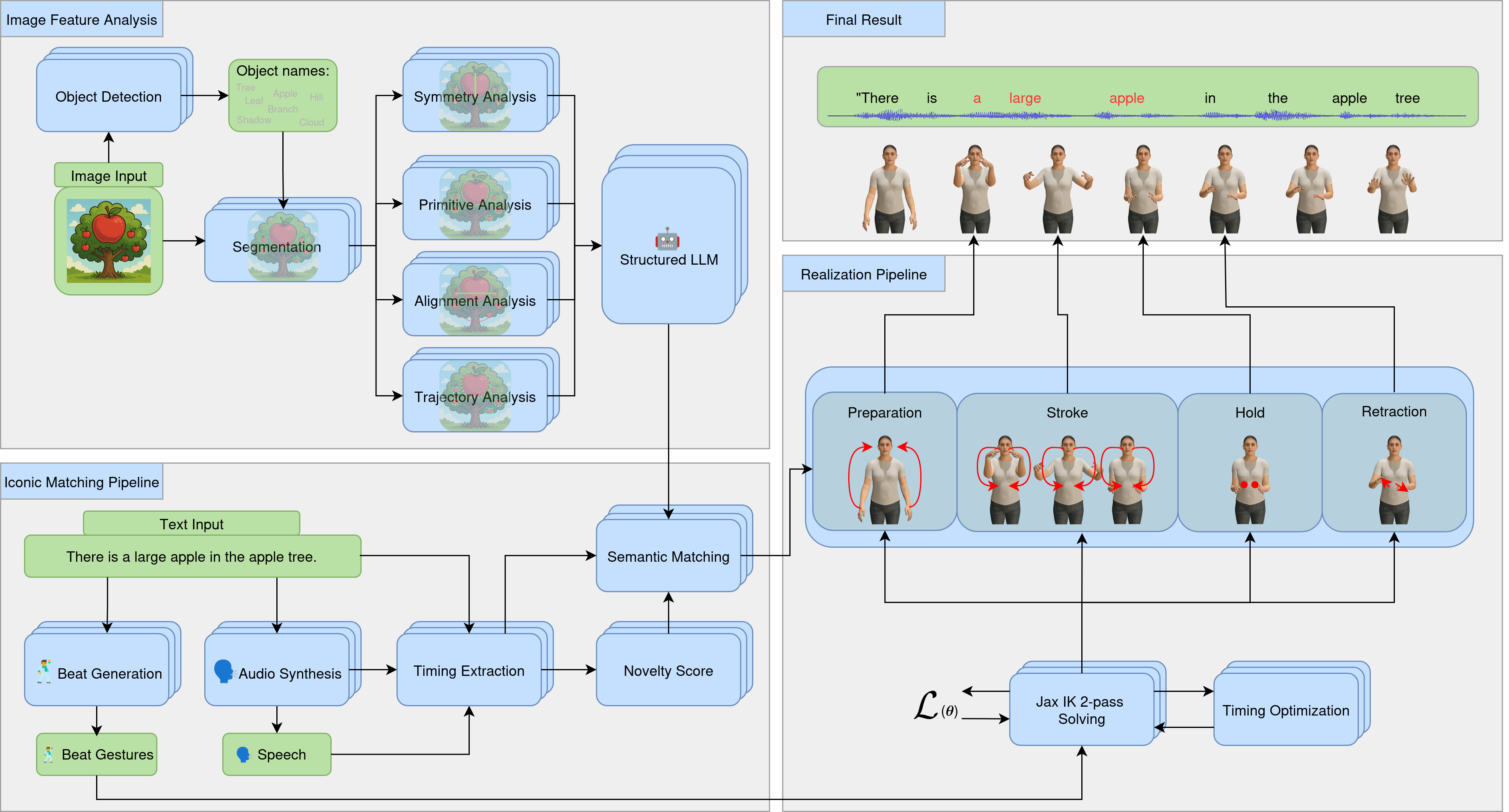}
    \caption{Architecture of the \modelplain{} model with its three main components:  Image Feature Analysis, Semantic Matching Pipeline, and Realization Engine}
    \Description{The diagram presents the \modelplain{} system designed to generate sign language gestures from both an image and its corresponding textual description. It is structured into four interconnected components: Image Feature Analysis, the Semantic Matching Pipeline, the Realization Engine, and the Final Result. The process begins with Image Feature Analysis, where an image is used as input. The system first performs object detection to identify visual elements like trees, apples, leaves, branches, hills, shadows, and clouds. Once these objects are detected, the image is segmented to isolate each component. Several types of analysis are then applied to the segmented image: symmetry analysis, primitive analysis, alignment analysis, and trajectory analysis. These analyses extract structural and spatial relationships between the objects in the image. The results are then passed to a structured large language model (LLM) that integrates this visual information with linguistic understanding. Parallel to this, the Semantic Matching Pipeline processes the text input, which in this case is the sentence "There is a large apple in the apple tree." The system generates beat gestures that emphasize rhythm and intonation, and performs audio synthesis to produce a spoken version of the sentence. Timing information is extracted from the speech, helping the system align gestures with audio. Semantic matching is then carried out to associate specific words in the sentence with relevant visual features extracted from the image. A novelty score is also computed to determine how uniquely the visual features match the linguistic concepts. These outputs feed into the Realization Engine, where the system plans and renders the gestures. Each gesture is divided into four stages: preparation (starting the movement), stroke (performing the core part of the gesture), hold (briefly maintaining a position), and retraction (returning to rest). The gestures are fine-tuned using a two-pass inverse kinematics solver, which ensures accurate and natural limb movements. Timing optimization ensures that the gestures are temporally aligned with the speech and meaning. Finally, the system produces a synchronized multimodal output. A digital avatar performs the generated sign language gestures in real time, matching the rhythm and content of the spoken sentence. The spoken text appears visually alongside the avatar, with each word highlighted as it is both spoken and signed. This final result offers an integrated and expressive communication experience that aligns visual, linguistic, and motor representations.}
    \label{fig:model_architecture}
\end{figure}

\section{Materials and methods}

The proposed model is structured into three main components (see Fig.~\ref{fig:model_architecture}): the Image Feature Analysis Pipeline, for extracting relevant objects from the image, the Semantic Matching Module, that matches the extracted objects to words in the given text, and the Realization Engine, that converts high-level concepts into iconic or deictic gestures. Each of these components addresses a specific stage in the process of generating co-speech iconic gestures, and together they enable an automatic end-to-end generation that can be integrated into any existing 3D co-speech gesture generation algorithm that has been trained on the SMPLX skeleton \cite{SMPL-X:2019}. This modular design ensures compatibility and flexibility, allowing for the seamless augmentation of gesture behaviors with semantically meaningful iconic gestures. In the following sections, we provide a detailed explanation of each component of the architecture.

A central design principle of our gesture generation pipeline is that the system's structure reflects the spatial requirements of iconic and deictic gestures. Iconic gestures often depict the form, size, and structure of a referent by translating visual properties, such as contours, symmetry, and geometric primitives, into hand and arm motions. Deictic gestures, by contrast, primarily encode the location and spatial arrangement of objects in a scene relative to the speaker or listener. Both types of gestures require precise spatial information about the referent: shape and contour information for iconic gestures, and size and relative position for deictic gestures. We structure our pipeline into separate modules for image feature extraction, semantic matching, and gesture realization to ensure that each stage addresses the distinct demands of these types of gestures. This allows the system to represent visual and spatial properties in a form suitable for generating human-like gestures.

This modular separation also reflects how gestures are naturally planned and executed. Iconic gestures typically trace contours or highlight salient features of objects. This necessitates a detailed analysis of object shapes before any motion can be generated. Similarly, deictic gestures require knowledge of the relative positions of objects and their alignment within a scene. This information must be derived from image-based spatial analysis rather than language alone. Capturing 2D spatial properties at the image analysis stage provides the downstream semantic matching and realization components with precise geometric and positional information. This allows the system to produce gestures that are semantically coherent and visually informative and ensures that the resulting gestures accurately encode the relevant aspects of the referent, while maintaining natural timing and motion characteristics consistent with co-speech behavior.

\subsection{Image Feature Analysis}
\begin{figure}[tb]
    \centering
    \includegraphics[width=1.0\linewidth]{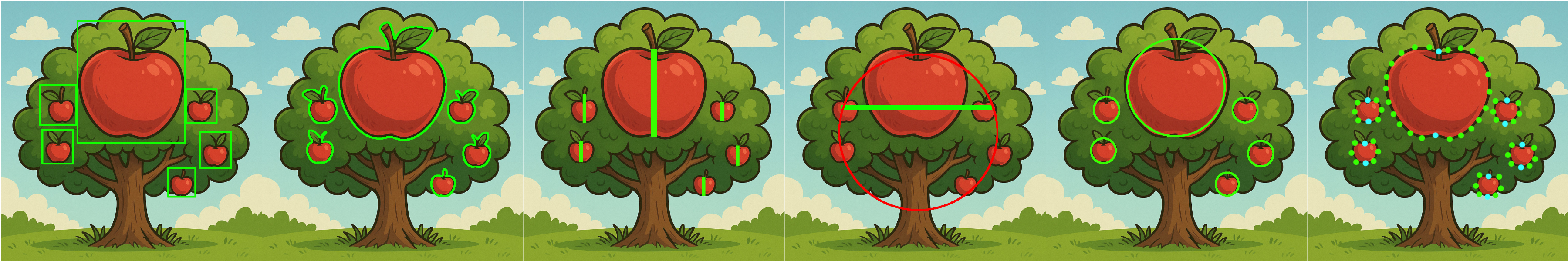}
    \caption{Illustration of all Image Feature Analysis steps on an example image. From left to right: Bounding Box detection, Segmentation, Object Symmetry Analysis, Object Alignment Analysis, Primitive Analysis, Trajectory Analysis.}
    \Description{}
    \label{fig:analysis}
\end{figure}

The goal of the Image Feature Analysis is to receive an image as input and to determine as much information about the objects in the image as possible. The rationale is that any or multiple of these objects may be referred to in the verbal utterances, and their visuospatial properties may hence be relevant for generating corresponding semantic gestures. As gestures often depict spatial aspects of a referent through forms or motion, we aim to extract not only object labels but also advanced structural properties, such as symmetry, geometric primitives, and alignment for each object. Different methods are used to detect, classify, and analyze the objects in the given image; a visualization of all steps is provided in Figure \ref{fig:analysis}.

As a first step, the input image is resized such that the longest edge measures 1024 pixels. During this process, the original aspect ratio of the image is preserved to prevent distortion. This resizing helps define certain thresholds on tiny items and normalizes the size of shapes in the image. Optionally, the background of the image can be detected and removed using the \textit{Rembg} library \cite{Gatis_rembg_2025}. This step helps isolate the main objects in the scene and can improve the accuracy of object detection and classification models by eliminating irrelevant visual context.

\paragraph{Object Detection}

To find all objects in the image, we prompt the vision-enabled Large Language Model \textit{Gemma 3} \cite{gemmateam2025gemma3technicalreport} six times using a diverse set of natural language prompts, each semantically similar to the phrase "What do you see in this image?" This multi-prompt approach is designed to elicit a broader and more comprehensive set of object predictions by encouraging the model to describe the scene from different linguistic perspectives.

To increase the diversity of the outputs and thereby improve object coverage in the scene, we use a relatively high temperature setting of 0.8 during inference. This introduces controlled randomness in the model's responses, allowing for a greater variety of plausible object descriptions across multiple runs. The resulting descriptive sentences are then lemmatized using the \textit{nltk} library \cite{bird-loper-2004-nltk}, which maps inflected words to their base forms. This normalization step ensures consistency in object naming and facilitates effective downstream filtering.

To isolate only physically present, nameable entities, we filter the lemmatized results against a curated list of 4555 concrete nouns. Concrete nouns, in this context, refer to terms that denote perceptible physical objects, such as "chair," "tree," or "car," as opposed to abstract concepts like "freedom" or "happiness."
Given that large language models are prone to hallucinations \cite{10.1145/3703155, Hong2024TheHL}, which is an even stronger problem in vision-enabled Large Language Models \cite{Zhong2024InvestigatingAM}, we introduce a verification stage to validate the presence of each proposed object. Specifically, for every object name extracted after filtering, we re-prompt Gemma 3 with a targeted verification prompt, asking whether that specific object is present in the image. The model is required to respond with a structured Boolean output (true or false) for each query, thereby confirming or rejecting the existence of the object in the image.

After obtaining verified object labels, we perform grounded segmentation to localize the identified entities within the image. For this task, we use \textit{Grounding DINO} \cite{liu2024groundingdinomarryingdino}, a model capable of zero-shot object detection without the need for fine-tuning. Grounding DINO takes the cleaned and verified labels as input and returns a set of bounding boxes along with associated confidence values, indicating the presence and location of each object in the scene. This step can also be viewed as an additional layer of verification, since only objects that are successfully localized by the detector are included in the final set. Following the bounding box detection, we apply the \textit{Segment Anything} Model 2 (SAM2) to generate pixel-wise masks for each detected object \cite{ravi2024sam2segmentimages}. SAM2 uses the bounding boxes produced by Grounding DINO as input prompts to refine the localization and produce high-resolution segmentation masks, completing the object detection.

\paragraph{Object Symmetry Analysis}

To assess the symmetry of each detected object, we first create a binary mask corresponding to the segmented regions. For each object, the image is then cropped to match the extent of this mask, ensuring that the symmetry analysis is focused solely on the object of interest. Next, the cropped image is rotated on its central vertical axis across a continuous range of angles from -90 to +90 degrees, with 1-degree increments. After each rotation, the image is bisected along the vertical axis, and the overlap between the two halves is computed. This overlap quantifies the degree of symmetry for each orientation. We record the overlap score at every rotation angle and identify the angle that yields the maximum overlap. This angle represents the orientation at which the object exhibits the highest degree of bilateral symmetry, providing a robust measure of visual symmetry aligned with the object's dominant axis. This symmetry information can be used in later steps to enable two-handed drawing gestures.

\paragraph{Object Alignment Analysis}

To evaluate the spatial alignment of objects belonging to the same class, we first extract the center coordinates of all bounding boxes associated with identical object labels. These center points serve as the basis for subsequent geometric analyses. This later enables us to either draw lines through objects corresponding to a specific alignment or discern different images based on their position.

We perform an exhaustive pairwise examination of all objects sharing the same label to detect possible linear alignments. Specifically, we assess whether three or more objects lie along a straight line within a tolerance of 5 degrees, accommodating slight deviations due to noise or perspective distortion. For each object, we identify the longest continuous sequence of aligned objects.

Once a group of aligned objects is found, those objects are removed from the current set, and the analysis is repeated on the remaining objects. This iterative approach continues until no further linear alignments can be detected or until no objects remain unassigned to an alignment group.

In cases where linear arrangements are absent, we further investigate whether five or more objects are positioned along a circular path. To this end, we collect the center points of potential candidates and estimate the circle’s radius. We then verify, within a predefined tolerance, whether all candidate points lie on the circumference of the estimated circle. As with linear alignments, any objects identified as part of a circular configuration are excluded from subsequent analyses, and the procedure is repeated with the remaining objects.

Finally, if after these analyses more than twenty objects remain without belonging to any linear or circular arrangement, we classify the residual objects as exhibiting a scattered or random spatial distribution. This assumption streamlines the analysis and prevents excessive computational effort in cases where no structured alignment is present.

\paragraph{Primitive Analysis}

As primitives are an essential part of iconic gestures and are easily detectable in human gesturing, we perform a primitive analysis. For each segmented object, we first crop the original image to the bounding region defined by its corresponding mask. The mask is then converted into a binary image to facilitate geometric analysis.

Using the image processing library \textit{OpenCV} \cite{opencv_library}, we systematically fit several primitive geometric shapes onto the binary mask. The primitives considered in this analysis include rectangles, triangles, circles, and ellipses. For each primitive type, a shape is fitted to the mask using a shape-specific fitting algorithm. Once a primitive is fitted, we calculate the overlap between the area of the fitted primitive and the original mask to assess how well the primitive represents the object’s shape.

The overlap is expressed as a percentage, indicating the proportion of the mask that coincides with the fitted primitive. For every object, we record the overlap percentages for all tested primitive shapes and identify the shape that yields the highest overlap. If the best-fitting primitive covers more than 85 percent of the mask area, we classify the object as resembling that respective primitive shape. This threshold ensures a high degree of geometric similarity and provides a reliable criterion for subsequent shape-based analyses in our pipeline.

\paragraph{Trajectory Analysis}

Two-handed drawing gestures are commonly performed as symmetrical gestures, with both hand trajectories starting at a joint upper point and following a mirrored movement into a joint lower end point\cite{kelso1984phase,byblow1999spontaneous,carson1990dynamics,dounskaia2010limitations}. To analyze whether any detected object would afford such a bi-manual drawing gesture, 
we first extract the motion paths from the masks corresponding to both hands. These paths serve as templates that both hands can subsequently follow during gesture execution, ensuring consistency in the modeled gestures.

To determine the starting point, the segmentation mask is first converted into a series of contour points. The contour is then simplified using the Ramer-Douglas-Peucker algorithm \cite{ramer1972iterative, douglas1973algorithms}, which preserves the essential shape of the object while reducing the number of points. This simplification helps to minimize computational complexity without compromising the fidelity of the contour. Following this simplification, we apply the Curvature Corrected Moving Average (CCMA) algorithm to smooth the contour and reduce noise \cite{Steinecker23}. The smoothed contour is used to compute the centroid of the object, defined by the average $x$ and $y$ coordinates of all contour points.

To locate the upper starting point, we select the point on the contour that has the highest $y$-value and lies closest to the calculated centroid in the $x$-dimension. Importantly, this point is not necessarily the global maximum in $y$, as it must also adhere to the object's geometric features. Once identified, we calculate the slope at this point using its adjacent neighbors on the contour. This slope $\theta$ can be expressed mathematically as:

$$
\theta = \tan^{-1} \left( \frac{y_{i+1} - y_{i-1}}{x_{i+1} - x_{i-1}} \right)
$$

Using this slope information, we iteratively traverse the contour in the upward direction, moving along the contour as long as the $y$-value continues to increase. This process terminates at the final cutting point, which serves as the topmost anchor for initiating gesture trajectories.

To generate trajectories for two-handed gestures, we introduce a novel coordination method termed the \textbf{SeeSaw} algorithm. This algorithm reflects the alternating nature of primary and secondary hand movement, where one hand leads and the other follows \cite{Quek2002MultimodalHD}. In this method, we designate a travel direction as the active one, typically corresponding to the dominant hand. Starting from the upper cutting point, the algorithm traverses the points of the contour in two directions: forwards (representing the dominant hand) and backwards (representing the non-dominant hand). The initial cutting point is added to both traversal lists to maintain symmetry.    

From this point, the active traversal list begins to grow. At each step, the next contour point is added to the active list and simultaneously removed from the original contour. After each addition, we evaluate whether the newly added point has crossed the $y$-position of the most recent point in the inactive list. If this condition is not met, the algorithm continues appending to the active list. Once the condition is satisfied, the active and inactive lists are swapped, and the process repeats. This bidirectional traversal continues until all contour points are exhausted.

The following pseudocode outlines the SeeSaw algorithm:

\begin{algorithm}
\caption{SeeSaw Algorithm for Two-Handed Trajectory Generation}
\begin{algorithmic}[1]
\Require Contour points $C$, initial cutting point $p_0$
\Ensure Left trajectory list $L$, right trajectory list $R$
\State Initialize $L \gets [p_0]$, $R \gets [p_0]$
\State Set \textit{active} $\gets L$, \textit{inactive} $\gets R$
\State Remove $p_0$ from $C$
\State Store initial $y$-direction sign: $\Delta y_{\text{sign}} \gets \text{sign}(y_{\text{next}} - y_{p_0})$
\While{$C$ is not empty}
    \State Take next point $p$ from $C$
    \State Append $p$ to \textit{active}
    \State Remove $p$ from $C$
    \State $y_{\text{last}} \gets$ y-position of last point in \textit{inactive}
    \State $y_{\text{current}} \gets$ y-position of $p$
    \If{sign($y_{\text{current}} - y_{\text{last}}$) $\neq \Delta y_{\text{sign}}$}
        \State Swap \textit{active} and \textit{inactive}
        \State Update $\Delta y_{\text{sign}} \gets \text{sign}(y_{\text{current}} - y_{\text{last}})$
    \EndIf
\EndWhile
\Return $L$, $R$
\end{algorithmic}
\end{algorithm}

This algorithm effectively creates a mirrored motion pattern, aligned on the y-axis of the contour for both hands, enabling synchronized gesture generation. That is, it enables dynamic trajectory extraction for bimanual gestures, preserving spatial and temporal coherence while adapting to the geometry of each object.

\subsection{Semantic Matching Module}

In the matching module, we are given a preprocessed image from the image pipeline and one or more natural language sentences that are expected to describe the scene or its contents. The primary objective of this module is to associate elements in the sentence with visual objects identified in the image.

To extract temporal information from the input text, we begin by synthesizing speech from the sentence using F5-TTS v1 \cite{chen-etal-2024-f5tts}, with a speech speed parameter set to 0.7. This slightly slowed-down synthesis allows for increased spacing between words, thereby enabling longer, more deliberate gestures to accompany each spoken element. Following synthesis, we extract word-level timestamps using the large English model version 3 of \textit{Whisper} \cite{radford2022whisper}, which provides fine-grained alignment between the audio and the textual representation.

For semantic grounding, we use the instruction-tuned \textit{Llama3.3} model with 70 billion parameters to match the information contained in the sentence with the preprocessed list of objects identified in the image \cite{grattafiori2024llama3herdmodels}. For this, we prompt the language model to return a structured output that includes several key elements. Specifically, the model must return the relevant phrase, preserving all adjectives and other descriptive modifiers. It must also identify the primary focus word of this phrase (for example, identifying the word "house" as the head noun in the phrase "a tall yellow house"). Additionally, the model is prompted to indicate whether any size information is mentioned in the phrase and, if so, to extract the specific size descriptor. Likewise, the output must specify any positional information and any alignment information referenced in the phrase.

If alignment data is detected within the structured output, the corresponding group of aligned objects is appended to the set of relevant objects. In cases where certain annotations (such as size or position) are not automatically detected by the model, we employ a manually constructed annotation dictionary. This dictionary defines canonical categories for size (e.g., tiny, small, medium, large, huge) and pre-defined deictic gesture points for position (e.g., left, right, bottom, top, top left, top right, middle left, middle right, bottom left, bottom right). The matching process uses an inverse-length matching strategy: longer phrases are matched first and removed from the search string to avoid overlapping annotations. For example, "bottom left" would be matched in its entirety, ensuring it is not subsequently split into individual matches for "bottom" and "left." These annotations are also augmented with explicit hand position and hand rotation specifications to support gesture generation.

Once object and phrase associations are established, we compute a Semantic Novelty score to assess the originality and communicative relevance of each phrase. This score determines whether a gesture should be performed for the phrase, based on the established view that gestures tend to accompany the parts of an utterance that contribute new information to the discourse \cite{McNeill1992HandMind}. The sentence is first parsed using \textit{spaCy} \cite{spaCy2020} to obtain token boundaries and sentence segmentation. Each phrase is then matched within its sentence, and a composite score is computed based on three components: semantic relevance, sentence position score, and a grammatical bonus.

Semantic relevance is calculated by embedding the phrase using \textit{MiniLM} \cite{wang2020minilm} and computing the maximum cosine similarity with all previously encountered phrase embeddings. The relevance score is then defined as one minus this maximum similarity:

\begin{equation}
\text{relevance}(p) = 1 - \max_{q \in \text{prev}} \cos\left( \text{emb}(p), \text{emb}(q) \right)
\end{equation}

The sentence position score captures the tendency for new information in English to appear toward the end of a sentence \cite{carlson2009information}. We assign a linear score between 0 and 1 depending on whether the phrase ends closer to the start or end of the sentence, with a score of 0 if it ends right after the first word and 1 if it ends after the final word. A binary bonus is also applied, with a value of 1 if the grammatical object of the sentence appears within the phrase, and 0 otherwise. These components are combined into a final information score using the following weighted sum:

\begin{equation}
\begin{aligned}
\text{importance}(p)  &= 0.65 \cdot \text{relevance}(p) + 0.25 \cdot \text{position}(p) \\
& + 0.1 \cdot \text{bonus}(p)
\end{aligned}
\end{equation}

Phrases that yield a novelty score below 0.5 are considered insufficiently informative and are excluded from triggering gesture behavior. This scoring mechanism ensures that only semantically important and contextually relevant information is visualized.

\subsection{Realization Engine}

To generate beat gestures corresponding to the input audio, we use the Expressive Masked Audio Gesture Modeling (EMAGE) algorithm \cite{liu2024emage}, which operates on the SMPLX skeleton. While EMAGE is the method employed in our current implementation, the modular nature of our pipeline allows for straightforward substitution with other co-speech gesture generation algorithms in future versions.

In order to focus the gesture synthesis on the upper body and hands, we zero out all joints that are not influenced during the deictic and iconic gesture generation. Specifically, only the collar, shoulder, elbow, wrist, and all finger bones of both hands are retained. Additionally, the root position of the skeleton is zeroed to normalize the global positioning of the avatar. This normalization helps with consistent spatial alignment. The zeroing transformation is preserved and can be reverted during post-processing, enabling seamless integration of beat and iconic gestures in the final animation.

An idle gesture is defined to serve as a default pose during periods of silence. This gesture represents the agent standing upright with both arms relaxed at the sides. To regulate the amplitude and expressivity of the generated beat gestures, we apply a blending strategy that combines the idle gesture with the EMAGE output. Specifically, we use a ratio of one part idle to three parts beat gesture. This weighted combination serves to suppress overly large or unnatural movements, especially those that extend above the head, while preserving the rhythmic and expressive qualities of the beat gestures.

To determine when the beat gestures should be displayed, we incorporate a Voice Activity Detection (VAD) module based on the Silero VAD model \cite{SileroVAD}. This is important as many beat gesture algorithms, including EMAGE, tend to generate large, over-exaggerated movements while the speaker is not speaking at all. This module operates at the frame level and determines whether the speaker is actively vocalizing. If speech is detected, the beat gesture is rendered in real-time. If no speech is detected, the system initiates a grace period of 0.5 seconds. Following this interval, the gesture smoothly transitions to the idle pose over the course of two seconds, creating a natural and fluid motion profile during pauses in speech.

For the deictic and iconic gestures, we define two primary types of prototype gesture movements for conveying semantic information: pointing gestures and trajectory gestures. Pointing gestures are designed to move a selected hand toward a specific target location in space, hold that position for a brief period, and then return to the resting pose. Trajectory gestures, in contrast, involve a more complex motion sequence. These gestures begin by moving the hand to the starting point of a predefined spatial path, following this path through a continuous motion, briefly holding the final position, and then returning to the neutral posture. Both pointing and trajectory gestures can be executed using either one hand or both hands, depending on the communicative intent and spatial characteristics of the referent.

Each of these gesture types can be further enriched with specific hand shape and orientation information. This includes configuring finger positions to represent a flat hand, a pointing finger, or a cupped shaping gesture, depending on the context. The direction of the hand's palm can also be specified to align with the spatial and semantic goals of the gesture, allowing for more expressive and accurate representations.

Building upon these motion and hand configuration templates, we implement four categories of automatically synthesized gestures: \textbf{single-handed trajectory gestures}, \textbf{two-handed trajectory gestures}, \textbf{abnormal bounding shape gestures}, and \textbf{alignment gestures}. Single-handed and two-handed trajectory gestures are used to follow one or more spatial paths, as derived from upstream modules or user specifications. Abnormal bounding shape gestures are triggered when the spatial dimensions of an object suggest a strongly elongated form. To detect such cases, we perform an Eigendecomposition on the covariance matrix of the object’s bounding box to identify the primary axis. We then compute the aspect ratio of this primary axis relative to its orthogonal counterpart. If this ratio exceeds a threshold of three to one, the system triggers a gesture specifically designed to communicate this disproportion.

Alignment gestures are used to represent the spatial alignment of objects based on predefined alignment data. These gestures follow either a linear or circular path, as calculated by our image processing pipeline, to indicate relationships such as collinearity or radial symmetry.

In addition to the gesture types described above, we implement two further variants of manual gestures:\textbf{ single-handed position gestures} and \textbf{two-handed size gestures}. Single-handed position gestures involve moving one hand to a specific location, as determined by the position information provided by the matching module. The hand moves to the target location, pauses briefly to hold the position, and then returns to its original position. This motion is used to emphasize or reference a particular point in space, providing clear spatial anchoring for the associated semantic content.

Two-handed size gestures, in contrast, are used to visually represent the dimensions of an object. In these gestures, both hands move into the space in front of the torso, oriented perpendicularly to each other with palms facing inward. The distance between the hands is determined by the size information provided, ranging from zero to fifty centimeters. After reaching the target distance, the hands hold this configuration momentarily before returning to their original positions. This gesture type effectively communicates the relative scale of the object in question.

All gesture trajectories are spatially normalized to ensure consistency across different scenes and avatar sizes. The normalization confines the motion paths within a bounding volume defined by the range x = [–50 cm, 50 cm], y = [–50 cm, 30 cm], and z = [30 cm, 50 cm], centered around the shoulder joint of the gesturing hand. This spatial constraint allows for predictable, scalable gesture animation while maintaining naturalness and clarity.

For all gesture types, whether manual or trajectory-based, we define a unified gesture execution model comprising the four major sequential gesture phases \cite{wagner2014}: Preparation, Stroke, Hold, and Retraction. Each phase is processed individually to allow for fine-grained temporal control and smooth transitions. The Preparation phase initiates the gesture by moving the hand from its current location to the first frame of the planned trajectory. The Stroke phase then follows the primary motion path associated with the gesture. Once the gesture reaches its intended endpoint, the Hold phase maintains the final position to provide visual emphasis and allow viewers to process the information being conveyed. Finally, the Retraction phase returns the hand to its resting position or prepares it for the next gesture.

To ensure temporal alignment between gesture and speech, we implement a gesture-speech alignment algorithm inspired by the empirical findings of \citet{bergmann2011relation}. Their research indicates that co-expressive speech and gesture (stroke) typically begin simultaneously, with a standard deviation of approximately 500 milliseconds. Based on this, our system synchronizes the stroke phase of each gesture with the primary focus word of the corresponding spoken phrase. This alignment enhances the naturalness and communicative effectiveness of the gesture. As it's possible that two gestures can overlap, we implement a priority stack to manage concurrent gestures. In this system, automatic gestures are given a higher priority, sorted by their novelty score, while manual gestures receive a lower priority. We solve the priority queue from the bottom up, so that higher priority gestures are solved last and can override lower priority gestures.

Automatic and manual gestures are constrained by minimum duration requirements to prevent abrupt or unnatural animations. Automatically synthesized gestures must have a minimum duration of three seconds, while manual gestures must last at least two seconds. This total time is distributed across the four gesture phases. For trajectory-based gestures, 20 percent of the frames are allocated to both the Preparation and Retraction phases, 5 percent to the Hold phase, and the remaining 55 percent to the Stroke phase. In the case of pointing gestures, which lack a stroke trajectory, the duration is instead divided such that Preparation and Hold each receive 40 percent of the frames, and Retraction receives 20 percent. These timing rules ensure that gestures are temporally balanced, expressive, and appropriate for real-time animation.

Two types of manual gestures were specifically designed to convey position and size information. Single-handed position gestures involve a single hand moving to a designated location based on position data obtained from the matching module. Once the target position is reached, the hand holds its location for a brief period before returning to its resting pose. This gesture is particularly effective for emphasizing specific spatial locations or points of interest within a scene.
The second type is two-handed size gestures, in which both hands move to a position in front of the torso, oriented perpendicular to one another, with palms facing each other. The distance between the hands is determined by the size information extracted from the input data. Depending on the magnitude of this size value, the hands stop at a separation ranging from 0 centimeters up to a maximum of 50 centimeters. After reaching this configuration, the hands hold the position momentarily before returning to their default states. This gesture type serves to visually represent the spatial extent or dimensional scale of an object in a highly intuitive way.

To solve the inverse kinematics for both deictic and iconic gestures, we make use of the recently published framework \textit{Jax-IK} \cite{Voss2025RealTimeIK}. This system employs differentiable functions for both forward and inverse kinematics, enabling the optimization of bone positions through the use of arbitrary objective functions. These objectives allow for precise control over the gesturing limb and enable the generation of smooth, realistic motion trajectories.

We apply four distinct objectives in our optimization process to synthesize natural and semantically correct gestures. The first is the distance objective, which minimizes the Euclidean distance between a target point and the current position of the end effector as defined by the forward kinematics transformation:

\begin{equation}
    \mathcal{L}_{\text{distance}}(\boldsymbol{\theta}) = \left\| \mathbf{t} - T_{\text{effector}}(\boldsymbol{\theta}) \right\|_2.
    \label{eq:distance_obj_repeat}
\end{equation}

The second is the known bone rotation objective, which reduces the deviation between a specified bone rotation and the current rotation. This objective is primarily used during the retraction phase of a gesture, where it ensures that the final frame aligns with subsequent gesture data:

\begin{equation}
    \mathcal{L}_{\text{known}}(\boldsymbol{\theta}) = \frac{1}{N} \sum_{i=1}^{N} \left\| \theta_i - \theta_i^* \right\|_2^2,
    \label{eq:known_obj}
\end{equation}
with \(N\) being the number of controlled degrees of freedom. 

The third is the derivative objective, which aims to minimize the velocity, acceleration, and jerk of the joint trajectories. By computing higher-order finite differences, this objective enforces smoothness throughout the entire gesture:

\begin{equation}
    \mathcal{L}_{\text{derivative}}^{(n)}(\boldsymbol{\theta}) = \sum_{t=1}^{T-n} \left\| \sum_{k=0}^{n} (-1)^k \binom{n}{k} \boldsymbol{\theta}_{t+n-k} \right\|_2^2,
    \label{eq:smoothness_general}
\end{equation}

where $n$ indicates the order of smoothness, such as velocity (n = 1), acceleration (n = 2), or jerk (n = 3), and $T$ denotes the number of trajectory points. This objective ensures that transitions between frames are gradual, reducing abrupt or unnatural movements in the generated gestures.

The fourth objective is the bone direction objective, which guides a bone, such as the wrist, to align with a specified direction. This is achieved by minimizing the angular difference between the current direction of the bone and a target direction vector:

\begin{equation}
    \mathcal{L}_{\text{direction}}(\boldsymbol{\theta}) = \left( \arccos \frac{\langle \mathbf{d}_b(\boldsymbol{\theta}), \mathbf{d}_t \rangle}{\|\mathbf{d}_b(\boldsymbol{\theta})\|\|\mathbf{d}_t\|} \right)^2,
    \label{eq:look_obj}
\end{equation}

All four objectives are combined into a single weighted cost function to control the balance among distance accuracy, pose matching, smoothness, and directional alignment:

\begin{equation}
    \begin{aligned}
        \mathcal{L}(\boldsymbol{\theta}) &=  \lambda_{\text{dist}} \mathcal{L}_{\text{distance}}(\boldsymbol{\theta}) 
        + \lambda_{\text{known}} \mathcal{L}_{\text{known}}(\boldsymbol{\theta}) \\
        & + \lambda_{\text{derivative}} \mathcal{L}_{\text{derivative}}(\boldsymbol{\theta})
         + \lambda_{\text{direction}} \mathcal{L}_{\text{direction}}(\boldsymbol{\theta}).
    \end{aligned}
    \label{eq:full_objective}
\end{equation}

Here, $\lambda_{\text{distance}}, \lambda_{\text{known}}, \lambda_{\text{derivative}}, \lambda_{\text{direction}}$ represent scalar weights that determine the relative influence of each objective on the final inverse kinematics solution. This approach allows for the generation of physically plausible and communicatively effective gestures across a wide variety of gesture types and motion patterns.

The optimization process begins by solving for the manually annotated gestures. Once the manually defined gestures are computed, the system proceeds to solve the automatically generated gestures. This sequential approach enables the automatic gestures to partially overwrite the manual ones if they temporally overlap. Such overwriting is intentional, as automatic gestures often encode more contextually relevant or informative motion patterns that enhance the overall expressiveness of the behavior.

\paragraph{Two-pass system}

To ensure that gesture speed is natural and consistent with human motion perception, we employ a two-pass system for gesture generation. The first pass involves solving the full gesture structure for each of the four gesture phases. Each segment contributes differently to the communicative function of the gesture, and accurate temporal modeling is essential to preserve its expressive characteristics. To account for natural timing variability, we define a percentage-based time range in which each gesture segment may be extended or shortened. Specifically, each segment is allowed to vary within 25\% of its original duration. Within this allowable range, we iterate through all possible new segment durations one frame at a time.

For each candidate segment duration within this time window, we perform the following operations. First, we compute the new number of frames for the segment based on the candidate duration. Then, we linearly stretch or compress the segment's motion to fit this new frame count. Next, we calculate the velocity, acceleration, and jerk profiles of the modified segment. To establish a reference for comparison, we extract the corresponding time window from the existing beat gestures and calculate its velocity, acceleration, and jerk over the same duration.

After computing the motion profiles for both the candidate and the reference beat gesture, we select the segment duration that minimizes the overall mean distance between them. This selection criterion is based on the similarity of kinematic profiles, ensuring that the final gesture not only maintains consistent speed but also aligns its beat structure with that of natural human gestures.

\begin{algorithmic}[1]
\Function{Timing Optimization}
{$\text{segment}, \text{beat\_ref}, \text{percent\_range}$}
    \State $T_{\text{orig}} \gets \text{length(segment)}$
    \State $T_{\text{min}} \gets T_{\text{orig}} \cdot (1 - \text{percent\_range})$
    \State $T_{\text{max}} \gets T_{\text{orig}} \cdot (1 + \text{percent\_range})$
    \State $\text{best\_score} \gets \infty$
    \State $\text{best\_segment} \gets \text{segment}$
    \For{$T_{\text{new}} \in [T_{\text{min}}, T_{\text{max}}]$}
        \State $\text{segment}_{\text{stretched}} \gets \text{LinearStretch}(\text{segment}, T_{\text{new}})$
        \State $(v, a, j) \gets \text{ComputeKinematics}(\text{segment}_{\text{stretched}})$
        \State $\text{beat}_{\text{window}} \gets \text{ExtractTimeWindow}(\text{beat\_ref}, T_{\text{new}})$
        \State $(v_b, a_b, j_b) \gets \text{ComputeKinematics}(\text{beat}_{\text{window}})$
        \State $\text{score} \gets \text{Diff}(v, v_b) + \text{Diff}(a, a_b) + \text{Diff}(j, j_b)$
        \If{$\text{score} $<$ \text{best\_score}$}
            \State $\text{best\_score} \gets \text{score}$
            \State $\text{best\_segment} \gets \text{segment}_{\text{stretched}}$
        \EndIf
    \EndFor
    \State \Return $\text{best\_segment}$
\EndFunction
\end{algorithmic}

\section{Results}

Previous studies to evaluate co-speech gesture synthesis have shown that most objective metrics have negligible correlation with human evaluation results and perform no better than chance, with the exception of the Fréchet Gesture Distance (FGD), which moderately correlates with human evaluation \cite{kucherenko2024evaluating}. However, since we don't have any human gesture data to compare our generated data against, in our case, it is not possible to use the FGD as an evaluation metric. Therefore, we focus on a user study to test our iconic and deictic gestures. This approach allows us to observe how well the gestures are perceived by human interlocutors and how much the performed iconic/deictic gestures influence the provided information.

\subsection{Human Rater-based Evaluation}

\begin{figure}[tb]
    \centering
    \includegraphics[width=0.8\linewidth]{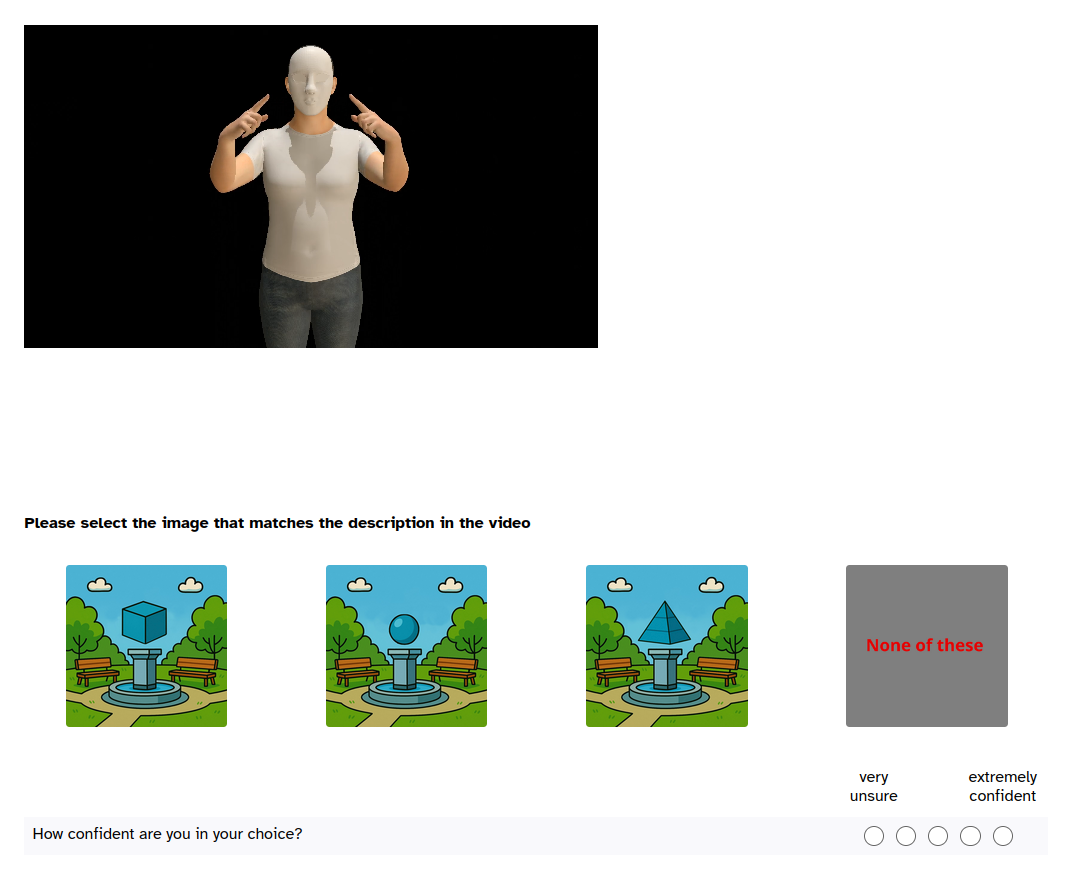}
    \caption{Example of the performed user study. First, the video has to be watched before selecting one of the four available images and a confidence value.}
    \Description{A webpage displaying a video of an SMPLX agent wearing a white shirt, arms slightly outstretched. Below the video is a prompt: 'Please select the image that matches the description in the video.' Three cartoon fountain images are shown: first with a blue cube on top, second with a blue sphere, and third with a blue pyramid. A fourth option says 'None of these' in red text on a gray background. Below is a confidence scale from 'very unsure' to 'extremely confident' with five empty circles for selection.}
    \label{fig:study_design}
\end{figure}

We used the SMPLX armature rendered in Blender for all visualizations (see Fig.~\ref{fig:study_design}). Since the EMAGE algorithm does not produce convincing lip synchronization, we applied a mask to the SMPLX skeleton. This approach ensures that participants are not distracted by the absence of facial animations or lip movement, allowing them to focus entirely on the gestures themselves.

The evaluation targeted five distinct categories of gestures. The first category included iconic gestures representing simple shapes such as rectangles, spheres, and triangles, called "Simple shape". The second category involved iconic gestures depicting more complex shapes, for example, the outline of a tree, called "Complex shape". The third category tested iconic gestures that conveyed alignment information, such as arrangements in a horizontal line, called "Alignment". The fourth category focused on iconic gestures representing size, such as indicating a very large tree or demonstrating a specific dimension, called "Size". Finally, the fifth category comprised deictic gestures that specified positional information, for instance, pointing to the left or referencing a particular side, called "Position".
    
The study used a 2×2 experimental design. The first factor was the presence or absence of semantically relevant information in the given text, and therefore, the spoken audio of the agent. The second factor was the type of gesture shown: either simple beat gestures or a combination of beat gestures with automatically generated deictic and iconic gestures. This design resulted in four conditions: text with all relevant information paired with beat gestures ("All Info + Beat"), text with all relevant information paired with both beat and our deictic/iconic gestures ("All Info + Iconic"), text with no relevant information paired with beat gestures ("No Info + Beat"), and text with no relevant information paired with both beat and our deictic/iconic gestures ("No Info + Iconic").

As an example of the difference in semantically relevant information, the given text could either be "The books in this bookcase are arranged horizontally" for all semantically relevant information or "The books in this bookcase are arranged like this" for no semantically relevant information. In all cases, we did not give our approach any additional annotations besides the provided text, which meant that the approach had to extract all relevant information, e.g., the "arranged like this" itself, and generate the correct gesture for this.  

Participants were asked to identify the correct object from three different images based on the information presented in the video. In addition to these three images, a fourth fallback option was always available to indicate that none of the displayed images were appropriate. In all cases, there was one correct image. An example page of the study design is shown in Figure \ref{fig:study_design}. For each of the five gesture categories, we prepared three different videos, resulting in a total of fifteen videos for the study.

After completing the main task, participants filled out an ASAQ survey \cite{FITRIANIE2025103482}, which measures factors such as Likeability, Performance, Believability, Coherence, Engagement, Enjoyability, and Acceptance of the virtual agent. This provides a comprehensive evaluation of the participants’ perception of the generated gestures and their overall interaction experience.

We recruited 105 participants for each condition, resulting in a balanced dataset. A performed power analysis indicated that a sample size of 80 participants per condition would be sufficient to detect a medium effect size (Cohen's f = 0.2) with a power of 0.8 and an alpha level of 0.05. To account for potential dropouts and ensure robust statistical power, we increased the sample size to 105 participants per condition, giving us a power of 0.943. This larger sample size also allows for more reliable detection of smaller effect sizes, enhancing the overall validity of our findings. All participants were native or fluent English speakers from the USA or UK, recruited through the Prolific platform \cite{prolific2024}. Screening ensured that participants had previously completed at least 20 surveys with a success rate of 95 percent or higher. We maintained gender balance and compensated participants at a rate of 10.5 euros per hour. The study was conducted using SoSciSurvey and followed a between-subjects design, where each participant was exposed to only one condition to avoid learning effects \cite{soscisurvey2024}. The average completion time was 14 minutes. We included four attention checks throughout the study. Two of these checks were embedded in the video tasks and required participants to select a specific image and report their confidence level. The other two checks were text-based and involved selecting specific answers to three separate questions each.
For the analysis, we included only participants who completed the entire survey and who did not fail any part of the four attention checks. Additionally, we removed participants whose completion time was either lower or higher than two standard deviations from the mean, as these outliers could indicate rushed or inattentive participation. After applying these criteria, we retained exactly 98 participants per condition for the final analysis, giving us 392 participants in total.

To assess the distribution of the data, we conducted Shapiro-Wilk normality tests for all conditions. The results indicated that none of the conditions met the assumption of normality. Furthermore, the assumption of homogeneity of variances was not satisfied in any condition. Due to these violations of parametric test assumptions, we used the Mann-Whitney-Wilcoxon test. In addition, to control for multiple comparisons and reduce the risk of Type I errors, we applied a Bonferroni correction to the p-values.

\subsection{Perception Study Results}

\begin{figure}[tb]
    \centering
    \includegraphics[width=1.0\linewidth]{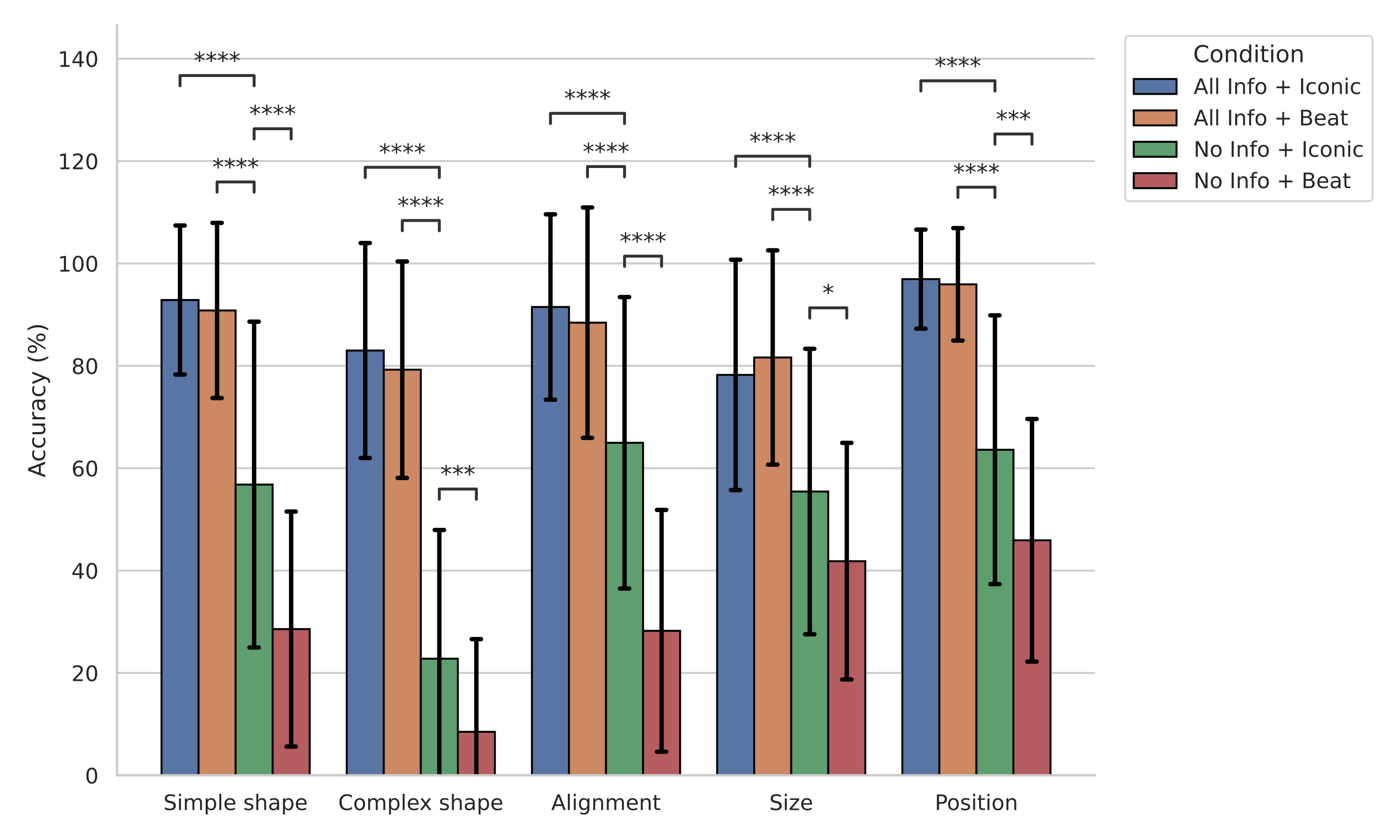}
    \caption{Mean accuracy for selecting the correct image from four options for each condition, grouped by the five tested categories with corresponding standard deviations. Statistical significance is only shown for the 'No Info + Iconic' condition. Higher values indicate better performance. ( * = p $<$ 0.05, ** = p $<$ 0.005, *** = p $<$ 0.0005, **** = p $<$ 0.00005)}
    \Description{Mean accuracy for selecting the correct image from four options for each condition, grouped by the five tested categories with corresponding standard deviations. Statistical significance is only shown for the 'No Info + Iconic' condition. Higher values indicate better performance. ( * = p $<$ 0.05, ** = p $<$ 0.005, *** = p $<$ 0.0005, **** = p $<$ 0.00005)}
    \label{fig:accuracy}
\end{figure}

The mean accuracy of choosing the correct image out of four possible images, for each condition and category, is shown in Figure \ref{fig:accuracy}. For visualization purposes, the figure shows only those significant differences that involve the condition "No Info + Iconic". This decision was made to enhance clarity and reduce visual complexity while still highlighting the most relevant contrasts for interpretation.

Overall, both "All Info" conditions achieved high accuracy across all categories, with means generally above 78\%. In contrast, the "No Info" conditions performed substantially worse, particularly for complex tasks.

For Alignment, accuracy was highest in "All Info + Iconic" (M = 91.50, SD = 18.11) and "All Info + Beat" (M = 88.44, SD = 22.50), whereas "No Info + Beat" dropped sharply to M = 28.23 (SD = 23.62). "No Info + Iconic" performed moderately (M = 64.97, SD = 28.47). All comparisons between "All Info" and "No Info" conditions were significant (p = 0.0000), and within the "No Info" conditions, "Iconic" outperformed "Beat" (p = 0.0000).

A similar pattern appeared for Complex Shape, where "All Info + Iconic" (M = 82.99, SD = 21.00) and "All Info + Beat" (M = 79.25, SD = 21.14) scored much higher than "No Info + Iconic" (M = 22.79, SD = 25.14) and "No Info + Beat" (M = 8.50, SD = 18.11). All pairwise differences between "All Info" and "No Info" conditions were highly significant (p = 0.0000), and "No Info + Iconic" was also significantly better than "No Info + Beat" (p = 0.0001).

For Position, both "All Info" conditions reached near ceiling accuracy (M = 96.94, SD = 9.68 for Iconic; M = 95.92, SD = 10.98 for Beat), whereas the "No Info" conditions were notably lower (M = 63.61 for Iconic; M = 45.92 for Beat). Differences between all "All Info" and "No Info" comparisons were significant (p = 0.0000), and "No Info + Iconic" again outperformed "No Info + Beat" (p = 0.0001).

In Simple Shape, "All Info + Iconic" (M = 92.86, SD = 14.56) and "All Info + Beat" (M = 90.82, SD = 17.11) were both substantially higher than "No Info + Iconic" (M = 56.80, SD = 31.82) and "No Info + Beat" (M = 28.57, SD = 22.95). All pairwise comparisons between "All Info" and "No Info" conditions were significant (p = 0.0000), and the two "No Info" conditions also differed significantly (p = 0.0000).

Finally, for Size, accuracy was lower overall compared to other categories, but the same pattern held. "All Info + Beat" reached M = 81.63 (SD = 20.93), slightly higher than "All Info + Iconic" (M = 78.23, SD = 22.50). "No Info + Iconic" (M = 55.44, SD = 27.89) and "No Info + Beat" (M = 41.84, SD = 23.11) lagged behind, with significant differences between all "All Info" and "No Info" comparisons (p = 0.0000) and a smaller yet significant advantage for "Iconic" over "Beat" in the "No Info" condition (p = 0.0136).

Across all categories, there was no significant difference between the two "All Info" conditions, whereas both "No Info" conditions performed significantly worse in every category. Notably, "No Info + Iconic" featured the highest variability across categories, with standard deviations reaching up to 31.82 in Simple Shape.

\begin{figure}[tb]
    \centering
    \includegraphics[width=1.0\linewidth]{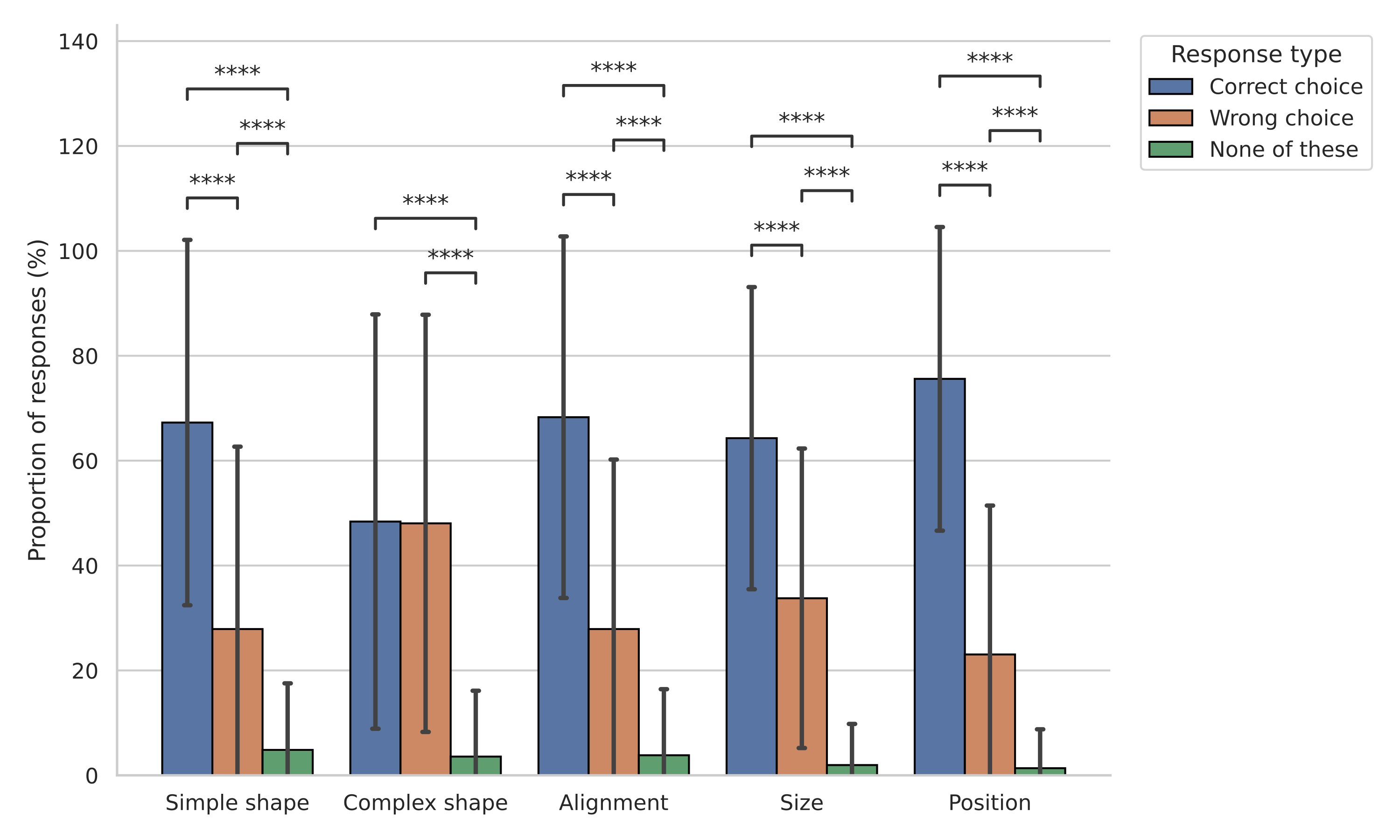}
    \caption{Proportional response rates for selecting the correct image, an incorrect image, or the fallback option "None of these", across all conditions, grouped by tested categories. ( * = p $<$ 0.05, ** = p $<$ 0.005, *** = p $<$ 0.0005, **** = p $<$ 0.00005)}
    \Description{Proportional response rates for selecting the correct image, an incorrect image, or the fallback option "None of these", across all conditions, grouped by tested categories. ( * = p $<$ 0.05, ** = p $<$ 0.005, *** = p $<$ 0.0005, **** = p $<$ 0.00005)}
    \label{fig:breakdown_all}
\end{figure}

The overall response rates for each category are presented in Figure \ref{fig:breakdown_all}. Across all categories, correct choices were selected significantly more often than incorrect ones (all p = 0.0000), with one key exception. In the "Complex shape" category, correct (M = 48.38, SD = 39.50) and incorrect choices (M = 48.04, SD = 39.77) were selected at nearly equal rates (p = 1.0000). In contrast, accuracy was highest for "Position" (M = 75.60, SD = 28.95), followed by "Alignment" (M = 68.28, SD = 34.47), "Simple shape" (M = 67.26, SD = 34.83), and "Size" (M = 64.29, SD = 28.81). Incorrect choices in these categories ranged from M = 23.04 (SD = 28.39) in "Position" to M = 33.76 (SD = 28.56) in "Size", and were consistently lower than correct choices. The fallback option, "None of these," was rarely selected across all categories, with means ranging from 1.36 to 4.85 and showing minimal variability. All pairwise comparisons involving "None of these" were also highly significant (all p = 0.0000).

\begin{figure}[tb]
    \centering
    \includegraphics[width=1.0\linewidth]{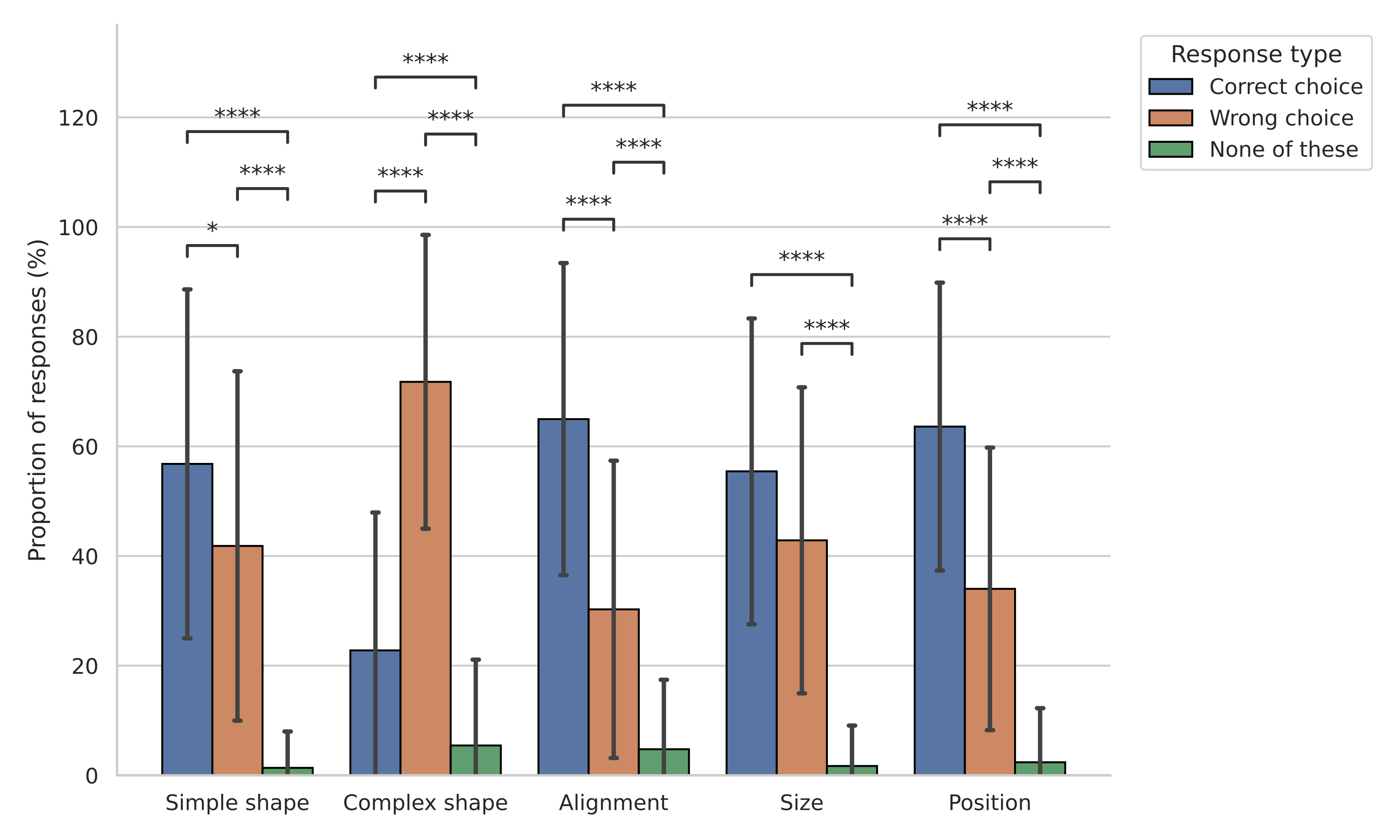}
    \caption{Proportional response rates for choosing the correct image, the wrong image, or the fall back option "None of these", for the condition "No Info + Iconic", grouped by the tested categories ( * = p $<$ 0.05, ** = p $<$ 0.005, *** = p $<$ 0.0005, **** = p $<$ 0.00005)}
    \Description{Proportional response rates for choosing the correct image, the wrong image, or the fall back option "None of these", for the condition "No Info + Iconic", grouped by the tested categories ( * = p $<$ 0.05, ** = p $<$ 0.005, *** = p $<$ 0.0005, **** = p $<$ 0.00005)}
    \label{fig:breakdown_none_iconic}
\end{figure}

Focusing on the "No Info + Iconic" condition (Figure \ref{fig:breakdown_none_iconic}), we observe a similar response pattern to the overall results, though with generally lower accuracy and reduced differences between response types. Correct choices still outnumbered incorrect ones in four out of five categories, with significant differences in each case (all p $<$ 0.05), except for "Size", where the comparison was marginal (p = 0.0690). Accuracy was highest in "Alignment" (M = 64.97, SD = 28.47) and "Position" (M = 63.61, SD = 26.25), closely mirroring the overall trends. However, correct response rates were noticeably lower for "Simple shape" (M = 56.80, SD = 31.82) and "Size" (M = 55.44, SD = 27.89), compared to their respective overall means of 67.26 and 64.29.

The most striking deviation from the overall pattern occurred in the "Complex shape" category. Here, the majority of responses were incorrect (M = 71.77, SD = 26.80), while correct choices dropped to just 22.79 (SD = 25.14), far below the overall mean of 48.38 and approaching chance performance. This inversion was highly significant (p = 0.0000), indicating particular difficulty in this condition when no contextual information was available. Across all categories, the fallback option "None of these" remained low, ranging from 1.36 to 5.44, with slightly elevated use in "Complex shape" (M = 5.44, SD = 15.65) compared to the rest. All pairwise comparisons involving the fallback option were significant (all p = 0.0000).

\begin{figure}[tb]
    \centering
    \includegraphics[width=1.0\linewidth]{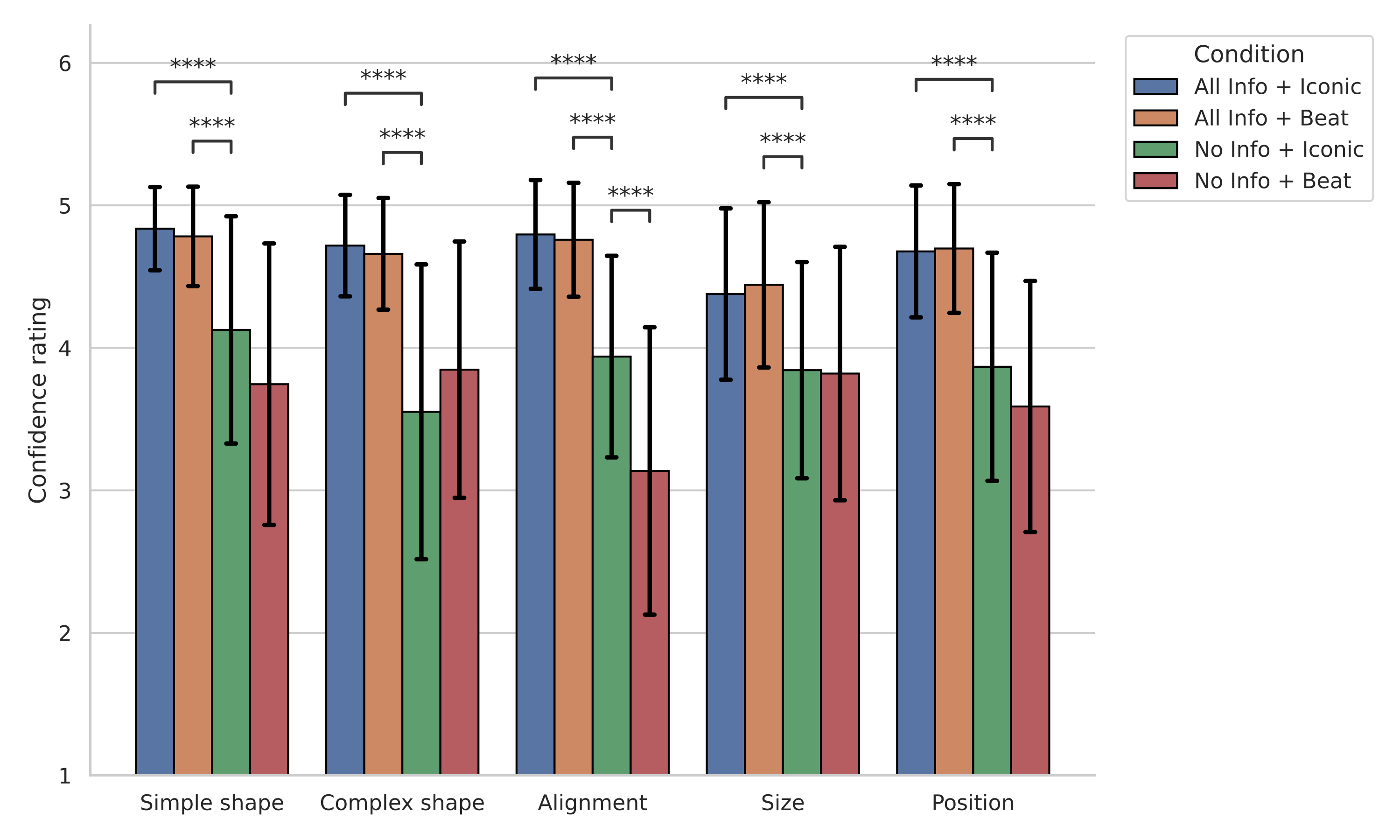}
    \caption{Mean self-reported confidence score for each condition, grouped by the five tested categories with corresponding standard deviations. Rated between 1 (very unsure) and 5 (extremely confident). Statistical significance is only shown for the 'No Info + Iconic' condition. Higher values indicate better performance. ( * = p $<$ 0.05, ** = p $<$ 0.005, *** = p $<$ 0.0005, **** = p $<$ 0.00005)}
    \Description{Mean self-reported confidence score for each condition, grouped by the five tested categories with corresponding standard deviations. Rated between 1 (very unsure) and 5 (extremely confident). Statistical significance is only shown for the 'No Info + Iconic' condition. Higher values indicate better performance. ( * = p $<$ 0.05, ** = p $<$ 0.005, *** = p $<$ 0.0005, **** = p $<$ 0.00005)}
    \label{fig:confidence}
\end{figure}

Figure \ref{fig:confidence} presents the mean self-reported confidence ratings for each condition. Consistent with previous findings, the "All Info" conditions produced uniformly high confidence scores, ranging from M = 4.38 (SD = 0.60) in "Size" to M = 4.84 (SD = 0.29) in "Simple shape." There were no significant differences between the two "All Info" conditions within any category. In contrast, the "No Info" conditions yielded substantially lower confidence ratings, with means ranging from 3.14 (SD = 1.01) to 4.13 (SD = 0.80). All comparisons between "All Info" and "No Info + Iconic" conditions were significant (all p = 0.0000).

Across four of the five categories, there were no significant differences between the two "No Info" conditions. The one clear exception was the "Alignment" category, where participants reported significantly higher confidence in the "No Info + Iconic" condition (M = 3.94, SD = 0.71) than in the "No Info + Beat" condition (M = 3.14, SD = 1.01; p = 0.0000). Smaller numerical differences between these two "No Info" conditions were also observed in other categories, such as "Simple shape" (M = 4.13 vs 3.74) and "Position" (M = 3.87 vs 3.59), but these did not reach statistical significance. Notably, the variance in confidence ratings was considerably higher in the "No Info" conditions compared to the "All Info" conditions.

\subsection{ASAQ Questionnaire Results}

\begin{figure}[tb]
    \centering
    \includegraphics[width=0.8\linewidth]{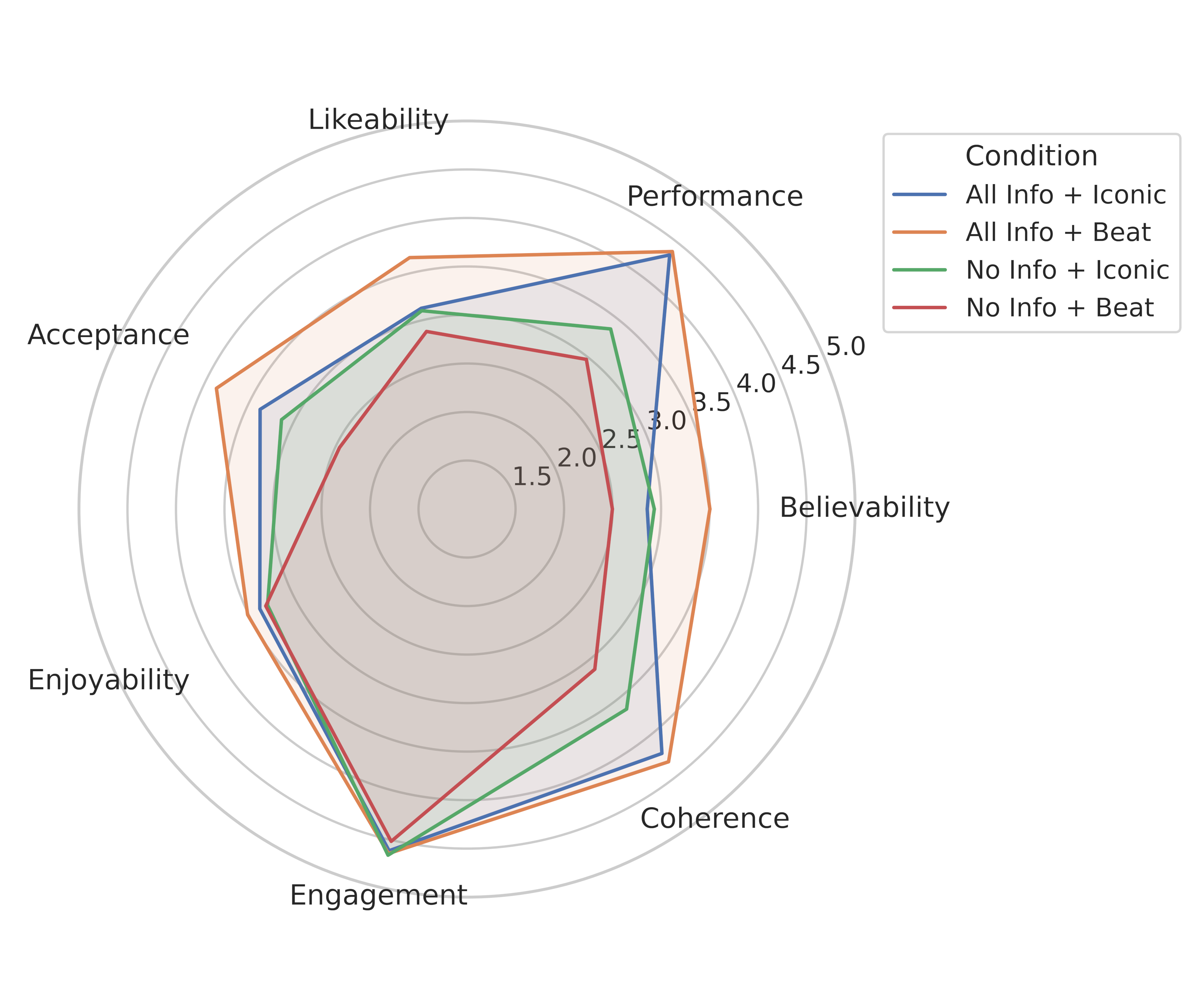}
    \caption{Mean scores of the ASAQ survey for all seven categories and conditions. Rated between 1 (disagree), 3 (neutral), and 5 (agree).}
    \Description{Mean scores of the ASAQ survey for all seven categories and conditions. Rated between 1 (disagree), 3 (neutral), and 5 (agree).}
    \label{fig:asaq_results}
\end{figure}

Figure \ref{fig:asaq_results} summarizes participants’ ratings across the seven ASAQ dimensions: Believability, Performance, Likeability, Acceptance, Enjoyability, Engagement, and Coherence, for each condition. Overall, responses revealed a strong effect of information availability, with the "All Info" conditions consistently receiving higher ratings than their "No Info" counterparts across most dimensions.

For Believability, ratings were highest in the "All Info + Beat" condition (M = 3.50, SD = 1.06), which was significantly higher than both "All Info + Iconic" (U = 5744.50, p = 0.0005) and the two "No Info" conditions. The lowest ratings occurred for "No Info + Beat" (M = 2.50, SD = 0.99), which differed significantly from "All Info + Beat" (p $<$ 0.0001).

In Performance, both "All Info" conditions scored substantially higher (M $\approx$ 4.36) than the "No Info" conditions, which dropped notably to M = 3.37 ("No Info + Iconic") and M = 2.97 ("No Info + Beat"). All pairwise comparisons between "All Info" and "No Info" conditions were significant (p $<$ 0.0001), confirming that performance perception is highly sensitive to informational support.

For Likeability, "All Info + Beat" again received the highest score (M = 3.66, SD = 1.11), significantly outperforming "All Info + Iconic" (p = 0.0045) and both "No Info" conditions (p $\leq$ 0.0063). Among the "No Info" conditions, ratings were lower overall, with the lowest for "No Info + Beat" (M = 2.88).

Acceptance showed a similar pattern, with "All Info + Beat" yielding the top rating (M = 3.87, SD = 1.26), significantly higher than all other conditions. "No Info + Beat" received the lowest acceptance score (M = 2.46) and differed significantly from "No Info + Iconic" (p = 0.0062).

For Enjoyability, differences were smaller across conditions (M $\approx$ 3.28–3.51). However, "All Info + Beat" was significantly higher than both "No Info" conditions (p $\leq$ 0.0115). No significant difference emerged between the two "All Info" conditions or between the two "No Info" conditions.

Engagement ratings were consistently high across all conditions (M $\approx$ 4.51–4.66) and did not differ significantly.

Finally, Coherence ratings favored "All Info" conditions (M = 4.22–4.33) over "No Info" conditions, which dropped to M = 3.64 ("No Info + Iconic") and M = 3.11 ("No Info + Beat"). All pairwise comparisons between "All Info" and "No Info" conditions were significant (p $\leq$ 0.0004). Within the "No Info" conditions, "Iconic" was rated more coherent than "Beat" (p = 0.0097).

\section{Discussion}

The primary objective of the present research was to address a significant shortcoming of speech-driven gesture synthesis and a challenge in virtual agent development by automating and scaling the generation of context-aware iconic and deictic gestures without relying on handcrafted annotations. Our comprehensive user study yielded evidence that our approach is viable and effective. It demonstrates that an agent can convey specific visual information through gestures alone, even when that information is absent from the accompanying speech.

Our system's success is most clearly demonstrated by its performance in the "No Info + Iconic" condition. Participants were significantly more accurate at identifying the correct object across all five gesture categories when the agent used the automatically generated iconic and deictic gestures compared to only simple beat gestures. The "No Info + Beat" condition serves as a crucial baseline, representing a scenario in which participants have no meaningful information from the ambiguous speech ("like this") or the nonrepresentational gestures. Accuracy in this condition hovered near or fell below chance levels in the case of Complex Shape, confirming that the task could not be solved by inference or bias in the experimental design alone. Therefore, the statistically significant increase in accuracy in the "No Info + Iconic" condition can be attributed to the information conveyed by the gestures. This finding validates our approach's core contribution to extract salient visual features from an image, align them with speech, and synthesize a gesture that communicates those features to a human observer in an informative and understandable way.

Furthermore, the results from the "All Info" conditions reveal another layer of nuance. The high accuracy in the "All Info + Beat" and "All Info + Iconic" conditions indicates that when speech is clear, the need for descriptive gestures to complete the primary task decreases. However, this does not diminish the value of the synthetic co-speech gesture as generated here. Rather, it highlights the potential of our system to create more robust and flexible conversational agents. Human speakers naturally adjust their use of gestures based on the assumed knowledge of their interlocutor and the complexity of the information being conveyed \cite{kuhlen2012gesturing, holler2020communicating}. Similarly, an agent equipped with our system could use explicit iconic gestures to clarify ambiguity, provide emphasis, or communicate with users in noisy environments where speech may be obscured. Conversely, it could resort to simpler beat gestures when the verbal channel is sufficient.

Closer examination of the results within each gesture category reveals important insights into the strengths and weaknesses of our system and the challenges inherent in gestural communication. The system successfully generated gestures for alignment and position. In the "No Info + Iconic" condition, these categories yielded the highest accuracy rates (65\% and 64\%, respectively). This is not surprising, as deictic and simple, linear iconic gestures are fundamental and highly common forms of nonverbal communication that clearly map to spatial concepts \cite{kopp2005spatial, alibali2005gesture}. The system's ability to identify object orientation and location and translate this information into a universally understood pointing or tracing gesture proves the effectiveness of the semantic matching module and the realization engine for core spatial information. The significant increase in self-reported confidence for "alignment" in the iconic condition compared to the beat condition suggests that participants found these gestures clearly communicative and helpful.

The Simple Shape category also performed well. Accuracy in the "No Info + Iconic" condition reached 57\%. These results demonstrate that the image extraction pipeline correctly identifies basic geometric shapes and that the gesture generation module can produce recognizable representations of them.

The size category posed a moderate challenge, with accuracy decreasing to 55\% in the "no info + iconic" condition. Although this was a significant improvement over the beat gesture condition, the lower performance and lack of a substantial confidence boost suggest an inherent ambiguity in communicating size through gestures. A gesture indicating "large" can be interpreted in multiple ways. Is the object large in height, width, or overall volume? Is it large relative to other objects in the scene, or is it large in an absolute sense? This ambiguity does not necessarily reflect a limit of the system, but rather a known complexity of human communication \cite{sassoon2011absolute, qing2014gradable, hassemer2016producing}. The fact that our system provided a significant informational boost despite this ambiguity is a positive outcome.

The most informative results came from the Complex Shape category. This category proved exceptionally difficult for participants in the "No Info + Iconic" condition. Accuracy plummeted to 23\%, and the response breakdown showed that incorrect choices were made much more frequently than correct ones. This result highlights a critical boundary condition for gestural communication. Translating an intricate, irregular shape, such as a tree, into a single, fluid hand trajectory poses immense difficulty for production and, conversely, for comprehension \cite{hassemer2018decoding, flanders2006planning}. The gesture may capture only a subset of the shape’s features, and the features chosen by the system may differ from those a human observer would consider most important for identification. These findings suggest that complex visual concepts may require more sophisticated gestural strategies, such as breaking shapes down into simpler components or using sequences of gestures \cite{hassemer2018decoding}. This presents a clear and valuable direction for future work.

The ASAQ survey results provide an interesting contrast to the objective accuracy data. Although iconic gestures were functionally superior at conveying information in ambiguous situations, they did not always lead to a better subjective experience. Notably, in the "All Info" conditions, the agent using simple beat gestures was rated significantly more believable, likable, and acceptable than the agent using iconic gestures. This apparent paradox can be understood through the lens of social robotics and human-agent interaction principles. When speech is clear and sufficient, as in the "All Info" conditions, highly descriptive, illustrative gestures can feel redundant and unnatural \cite{yeo2017teachers, yeo2017evidence}. They may make the agent appear overly literal or theatrical, which violates human conversational norms \cite{robrecht2024study}. In this context, the simpler, less semantically loaded beat gestures may serve as a more natural and unobtrusive form of conversational accompaniment, leading to a more positive subjective rating.

However, the coherence results tell a different story that strongly supports the presented approach. In the "No Info" conditions, the agent using iconic gestures was rated as significantly more coherent and acceptable than the agent using beat gestures. This is a critical finding. Even when the task was difficult and participants were not always accurate, as in the "Complex Shape" condition, they perceived the agent's attempt to communicate through meaningful gestures as more coherent than an agent that provided no visual cues. These results suggest that the use of iconic gestures aligns with user expectations of how a helpful agent should behave in situations of linguistic ambiguity.

Finally, the consistently high Engagement scores across all four conditions suggest that the experimental setup was compelling and that the agent, regardless of gesture type, was able to maintain user attention. The high-performance ratings for both "All Info" conditions confirm that users recognized when the agent was successfully completing its communicative goal.

\section{Limitations}

Although the results demonstrate a significant advancement in the automated generation of meaningful gestures, they also reveal several limitations of the system and outline a clear path for future work. These challenges underscore the need for further innovation to develop more seamless, intelligent, and human-like conversational agents.

A primary constraint of our current system is its inability to operate in real-time. The computational pipeline is currently too time-intensive for fluid, responsive interaction. On average, the image feature analysis pipeline takes 112 seconds to extract salient features, followed by an average of eight seconds for the matching module. Additionally, the realization engine, which translates desired motions into joint movements for the agent, takes approximately 20 seconds for each gesture. These delays are largely due to the current brute-force approach in the two-pass solving method. For an agent to be truly conversational, it must perceive, process, and respond incrementally and within the fluid turn-taking structure of human dialogue. This capability still requires substantial optimization of the framework presented here.

Another significant limitation, identified in the user study, is the system's difficulty in rendering complex shapes. The accuracy of participant responses for the "Complex Shape" category was notably low, revealing a fundamental challenge in gestural communication. Our system currently generates a gesture by tracing the external contour of an object. This approach is insufficient for intricate or irregular shapes, like a tree, where internal details or structural components are more defining than the overall silhouette. The act of translating a complex 2D image into a comprehensible 3D hand trajectory is an immense challenge, as the system may select features that a human observer does not find salient for identification. To our knowledge, no effective algorithm for the decomposition of general 2D images into semantically meaningful parts for gesturing is currently available. While recent developments in 2D to 3D image decomposition are promising, they are not yet robust enough for general application \cite{lin2025partcrafter}. Future work must explore more sophisticated strategies, such as breaking down complex objects into a sequence of simpler salient components. Research on human communication can provide important hints as to how complex information is being portioned and packaged such that it becomes readily conveyable in both speech and gesture \cite{slobin1987thinking}; cf.~\cite{wagner2014}.

Finally, our system does not yet adequately model the social nuances of communicative redundancy. As revealed in the "All Info" conditions of our study, the iconic gestures were perceived as less believable and likable when speech was perfectly clear. This suggests that the gestures, though technically accurate, became redundant and conversationally inappropriate. The presented model can easily remove semantically redundant gestures but does not yet reason about informational redundancy from the listener's perspective. It lacks the crucial social intelligence to determine when a simple beat gesture is more appropriate than a detailed iconic one. This points to a rich area for future research: developing partner models that can assess the clarity of the verbal channel and the user's likely state of knowledge to dynamically select the most appropriate level of gestural detail, thereby creating a more natural and socially aware conversational partner.

\section{Conclusion}

This paper presented an approach to address a critical limitation in co-speech gesture generation: existing systems operate solely on verbal input and cannot automatically produce expressive and context-aware semantic (iconic and deictic) gestures. We present a novel, zero-shot system that operates on multimodal (speech and image) inputs and successfully closes this gap. The system generates synchronized gestures that convey specific visual information derived from images and given text without requiring manual annotation or intervention. It does so by extending current speech-driven gesture synthesis architectures that are based on generative network models, by integrating a sophisticated image analysis pipeline with a semantic matching module and an inverse kinematics-based realization approach.

A comprehensive user study validated the approach and demonstrated that this system can generate interpretable and functionally effective gestures. In situations of linguistic ambiguity, where speech alone was insufficient, the generated iconic and deictic gestures significantly improved participants' ability to identify the correct properties of objects, such as alignment, position, and shape. These results confirm that the system can successfully extract salient visual features from the image input, map them to appropriate speech segments, and translate them into meaningful gestures for human users. The fact that these gestures were perceived as significantly more coherent in ambiguous contexts underscores their value in creating more robust, collaborative virtual agents.

At the same time, the evaluation revealed important nuances and future challenges. Participants' difficulty with complex shapes highlights a key boundary condition for single-stroke gestural representation. It also points to the need for more sophisticated strategies, such as breaking down complex shapes into sequences of gestures and multimodal utterances more generally. Additionally, subjective ratings from the ASAQ survey offer crucial insights. While information-rich gestures are vital for clarity in ambiguous contexts, they may be perceived as unnatural or unlikable when speech is clear. This suggests that believable agent interaction in the future will not simply involve generating more complex gestures, but rather developing models that can intelligently decide when and how to use them in conjunction with appropriate verbal behavior.

Overall, the present work demonstrates the feasibility of a fully automated process for creating iconic and deictic gestures that improve communication. Building on these findings, future work will refine the representation of complex concepts and imbue agents with the social intelligence to adapt their nonverbal behavior dynamically to the needs of the conversation. This will pave the way for more natural, effective, and truly multimodal communication between humans and embodied agents.

\clearpage
\bibliographystyle{ACM-Reference-Format}
\bibliography{references}

\end{document}